\documentclass[%
 reprint,
 amsmath,amssymb,
 aps,
]{revtex4-2}

\usepackage{graphicx}
\usepackage{dcolumn}
\usepackage{bm}
\usepackage{amsmath}
\usepackage{amssymb}
\usepackage{graphicx}
\usepackage{subfigure}
\usepackage{braket}
\usepackage{harpoon}
\usepackage{tikz}
\usepackage{comment}
   
\begin{document}

\preprint{APS/123-QED}

\title{Enhanced Valley Splitting in Si Layers with Oscillatory Ge Concentration}

\author{Yi Feng}
\affiliation{Physics Department, University of Wisconsin-Madison, 1150 University Ave, Madison, WI, 53706, USA}

\author{Robert Joynt}
\affiliation{Physics Department, University of Wisconsin-Madison, 1150 University Ave, Madison, WI, 53706, USA}

\date{\today}

\begin{abstract}
The valley degeneracy in Si qubit devices presents problems for their use in quantum information processing.  It is possible to lift this degeneracy by using the Wiggle Well architecture, in which an oscillatory Ge concentration couples the valleys. This paper presents the basic theory of this phenomenon together with model calculations using the empirical pseudopotential theory to obtain the overall magnitude of this effect and its dependence on the wavelength of the concentration oscillations.  We derive an important selection rule which can limit the effectiveness of the Wiggle Well in certain circumstances.    
\end{abstract}

\maketitle


\section{Introduction}
\label{sec:introduction}

Silicon-based spin qubits enjoy many advantages for quantum computing devices  \cite{Zwanenburg:2013p961}. They have longer intrinsic spin coherence times due to weak spin-orbit coupling.  There is also the possibility of eliminating decoherence from coupling to nuclear spin because of the existence of an abundant spin zero isotope.  Scaling to many qubits presents difficulties in all quantum computing platforms, but for Si there is at least a technological infrastructure already in existence for related purposes.  

One disadvantage of Si is the presence of the valley degree of freedom, a source of leakage of quantum information. This creates a degeneracy that is sample-dependent  and notoriously difficult to control. The degeneracy is split in real devices and the energy difference is referred to as the valley splitting (VS).  The barriers that confine the electrons to the active Si layer are known to do this, but the VS is sensitive to the details of the barrier.  As a result,  the VS is experimentally highly variable.  It ranges roughly from 30 to 250 $\mu$eV in SiGe/Si/SiGe  structures \citep{Weitz:1996p542, Koester:1996p1400, Lai:2006p161301, Goswami:2007p41, Mi:2015p035304,   MiPRL2017, Neyens:2018p243107}, while in MOS structures it tends to be considerably larger but still quite variable, with values ranging from 300 $\mu$eV up to nearly 1 meV \citep{Yang:2013p3069, Gamble:2016p253101}.  Overall, the barrier effects on the VS are reasonably well understood theoretically, a major theme being that a strong electric field perpendicular to the Si layer can push the wavefunction up against the interface, which tends to increase the VS  \citep{Boykin:2004p115, Boykin:2004p165325}. 

An important goal of research in this field is to somehow control the VS so that it is reliably larger than 200 $\mu$eV. One recent approach is to insert an ultra-thin layer of SiGe in a Si/SiGe heterostructure, which increases VS by about a factor of two \cite{mcjunkin2021spike}.  Adding Ge at random positions in the Si layer is also effective \cite{wuetz}.  
One may also add Ge to the Si layer in such a way that the Ge concentration has an oscillatory profile in the direction perpendicular to the layer \cite{mcjunkin2021wigglewell}.  This is called the Wiggle Well (WW) architecture.  It was shown that the added Ge lowered the mobility of the structure but that this did not preclude efficient device operation.

In this paper we describe in detail the basic ideas behind the WW and we present calculations of the VS under various conditions.  The calculations support the conclusion that splittings can be engineered to lie in the $5-15$ meV range, well above the values needed needed to eliminate leakage during qubit operation. We also derive a selection rule that strongly affects the VS in the device used in Ref.~\cite{mcjunkin2021wigglewell}.

The physical basis of the WW is described in Sec.~\ref{sec:calculation}, and the details of our  computational method in Sec.~\ref{sec:comp}.  The selection rule is proved in Sec.~\ref{sec:selection}.  The results are given in Sec.~\ref{sec:res}. Sec.~\ref{sec:conclusion} contains further discussion and a conclusion.

\section{Wiggle Well}
\label{sec:calculation}

\subsection{Valley Structure}
\label{subsec:physical}

Silicon is an indirect bandgap semiconductor with a valence band maximum at $\mathbf{k}=(0,0,0)$, and it has six degenerate conduction band minima along the $(001)$ and equivalent directions. This paper concerns the electron states in the Si layer of a SiGe/Si/SiGe heterostructure or in a MOS structure.  In the Si layer there is strain or other anisotropies present that reduce the degeneracy of the conduction band minimum to two \citep{Ando:1982p437}, at the k-points $\pm\mathbf{k} = \pm (0,0,k_0)$ with $k_0 = 0.84 (2\pi / a) $ where $a=0.543$ nm is the lattice constant of Si.  The z direction is perpendicular to the plane of the layer.  

\subsection{Hamiltonian}

We take the electrons in our model to be confined to a Si-rich layer. We shall deal with a two-dimensional electron gas that has translational invariance in the $x-y$ plane and apply periodic boundary conditions in these directions.    
The total Hamiltonian is
\begin{equation}
H_{tot} = H_{cr} + V_{tot}(z) =  H_{cr} +V_{str}(z) + V_d(\mathbf{r}) + V_{osc}(z).
\end{equation}
Here $H_{cr}$ is the unperturbed bulk Si Hamiltonian. $H_{cr}$ could also include the effects of strain, particularly if we are dealing with a SiGe/Si/SiGe system, but for simplicity we assume no strain in this paper.  This allows us to focus on the effects of the oscillatory potential $V_{osc}(z)$. $V_d(\mathbf{r})$ is the atomistic disorder potential produced by the Ge atoms in the well.  We will comment on this below, but again it is not the main focus. $V_{str}(z)$ denotes the device structure potential, which we take to have the form 

\begin{equation}
\label{eq:str}
V_{str}(z) = V_{b0} [1+\tanh(z/w)]/2 -eFz/\epsilon    
\end{equation}.  

The first term is a sharp step-like barrier potential and the second represents electric potential from an applied electric field $F$ and a dielectric constant $\epsilon$.  We use $V_{b0} = 1\text{eV}$ (a value more typical for MOS structures), $w= 1\text{nm}$, and $\epsilon=11$. The electric field along the z direction keeps the electrons close to the interface.  $V_d $ is the disorder potential, nonzero because the added Ge in the Si layer is not fully ordered. 

$V_{osc}$ stands for the oscillating potential.  The virtual crystal approximation for $V_{osc}(z)$ is 
\begin{equation}
\begin{aligned}
V_{osc}(z) = V_0 \, \overline{n}_{Ge} (1+\cos(qz)) 
\end{aligned}
\end{equation}
where $\overline{n}_{Ge}$ is the average fractional concentration of Ge in the predominantly Si layer. We take $V_0$ = - 0.5 eV. This is the value that gives the measured change in the energy of the conduction band minimum in the regime of low Ge concentration in strained Si$_{1-x}$Ge$_x$ layers \cite{schaeffler1997}.    

\begin{figure}[h!] 
    \centering
    \includegraphics[width=0.45\textwidth]{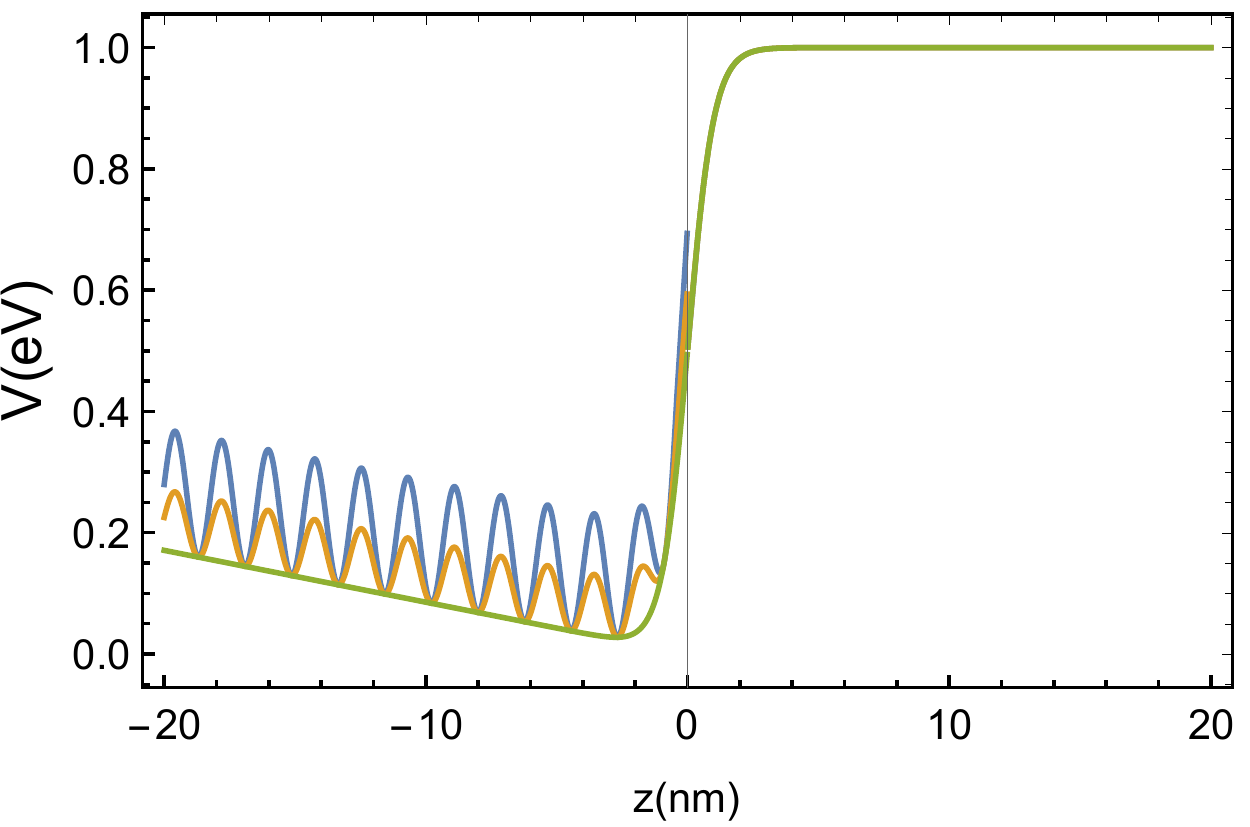}
    \caption{The smooth potential $V(z) = V_{str}(z) + V_{osc}(z)$ felt by an electron in the proposed SiGe heterostructure, shown for an average Ge concentration in the well of 0\%, 10\%, and 20\%, and an electric field $F/\epsilon = 8.5 \text {mV}/ \text{nm}$.  The atomistic disorder potential $V_{d}(\mathbf{r})$ is not visible on this relatively coarse scale. The Si layer with Ge added in a modulated fashion occupies the half space $z<0$. The barrier occupies the $z>0$ region.  The width of the barrier is 1 nm and its height is 1 eV.}
    \label{fig:potential}
\end{figure}

We give a sketch of the potential $V_{str}(z) + V_{osc}(z)$ for various Ge concentrations in Fig.~\ref{fig:potential}. 

$V_{osc}(z)$ is the defining feature of the WW.  It is created when the structure is grown by  depositing Ge atoms in a  sinusoidal fashion.  The effect of $V_{osc}(z)$ is to enhance the valley splitting, as will be explained in the next section. 

\subsection{Perturbative Picture}
\label{subsec:pert}

In this section we will treat the valley splitting in first-order perturbation theory.  It will be seen later that this is not sufficient for accurate calculations.  The reason for presenting it here is that it allows us to illustrate the physical ideas behind the WW and to determine the best candidate wavelengths for the modulation of the Ge concentration in the Si layer.

The zero-order Hamiltonian $H_{cr}$ is that of the pure bulk material. In the conduction band we have the Schrödinger equation
\begin{equation}
\label{eq:crystal}
    H_{cr}\psi_{k}(\mathbf{r})=H_{cr}[u_{\mathbf{k}}(\mathbf{r}) e^{i\mathbf{k} \cdot \mathbf{r} }]
    =\varepsilon(\mathbf{k}) \psi_{\mathbf{k}}(\mathbf{r}),
\end{equation}
that is, $u_{\mathbf{k}}(\mathbf{r}) e^{i\mathbf{k} \cdot \mathbf{r}}$ are the eigenfunctions of $H_{cr}$, the Hamiltonian of the silicon crystal.  $u_{\mathbf{k}}(\mathbf{r})$ is the lattice-periodic part of the Bloch function. $\varepsilon(\mathbf{k})$ is the band energy of an electron in the pure bulk system.  Our interest is when $\mathbf{k}$ is near one of the minima of the conduction band $\pm\mathbf{k}_{0}$, so $\varepsilon_{(\pm)}(\mathbf{k}) = \hbar^2 (k_{x}^2+k_{y}^2)/m_{t} +\hbar^2(k_{z} \pm k_0)^2/m_z$.  $m_t=0.92m_e$ is the transverse mass and $m_z=0.19 m_e$ is the longitudinal mass, where $m_e$ is the bare mass.  Since $u_{\mathbf{k}}(\mathbf{r})$ is periodic in the direct fcc lattice it has a Fourier expansion
\begin{equation}
\label{eq:ck}
    u_{\mathbf{k}}(\mathbf{r}) = \sum_{\mathbf{K}} c(\mathbf{K},\mathbf{k}) e^{i\mathbf{K} \cdot \mathbf{r} }, 
\end{equation}
where $\mathbf{K}$ runs over the bcc reciprocal lattice. $c(\mathbf{K},\mathbf{k})$ depends on $\mathbf{k}$ in general, but we will only need it when $\mathbf{k} \approx \pm \mathbf{k}_{0}$.  Thus we define $c_{\pm}(\mathbf{K}) = c(\mathbf{K}, \pm \mathbf{k}_0)$, and later assume that $c(\mathbf{K})$  is not a rapidly varying function of $\mathbf{k}$ near $\pm\mathbf{k}_{0}$.  Note that $c_{\pm}(\mathbf{K})=c_{\mp}^{*}(-\mathbf{K})$.   
Saraiva \textit{et al.} have used density functional theory to calculate the $c_{\pm}(\mathbf{K})$ in pure bulk Si~\cite{saraiva} which provides a good benchmark for our work. The $c_{\pm}(\mathbf{K})$ are modified by the presence of the added Ge in the Si layer.  This turns out to be an important effect, and we will discuss the computation of the  $c_{\pm}(\mathbf{K})$ in detail below.  
    
The states at the conduction band minima are 
$\psi_{\pm\mathbf{k}_0}(\mathbf{r})$
and satisfy
\begin{equation}
H_{cr}  \psi_{\pm\mathbf{k}_0}(\mathbf{r}) = \varepsilon_0 \psi_{\pm\mathbf{k}_0}(\mathbf{r}).
\end{equation}
The two wavefunctions $\psi_{\pm\mathbf{k}_0}(\mathbf{r})$ are degenerate.   
    

In first-order perturbation theory the total valley splitting $\Delta $ is \cite {goswami, saraiva}:

\begin{align}
\label{eq:formula}
\Delta =& 2 \, | \bra{\psi_{+\mathbf{k}_0}(\mathbf{r})} V_{tot} \ket{\psi_{-\mathbf{k}_0}(\mathbf{r})} | \\ 
=& 2 \, \big|\sum_{\mathbf{K},\mathbf{K^{'}}} c^*_+(\mathbf{K}) c_-(\mathbf {K^{'}})
 \delta_{K_{x},K_{x^{'}}} \delta_{K_{y},K_{y^{'}}}  I(K_{z} - K_{z^{'}})\big|    
\end{align}

where the last factor stands for the integral

\begin{equation}
\label{eq:integral}
I(K_z - K_{z}^{'})=\int_{-\infty}^{\infty} e^{iQz} V(z) dz 
\end{equation}
with $Q=K_z-K_z^{'}-2k_0$.  

The total valley splitting in the current approximation is:

\begin{equation}
\label{eq:splitting}
\Delta = |\Delta_w+\Delta_b+\Delta_d|.
\end{equation}
Here $\Delta_b$ is the barrier contribution. $\Delta_d$ is the disorder contribution which has been calculated recently \cite {wuetz}. 
$\Delta_w$ is the WW contribution, which is the subject of this paper.  It is caused by the oscillatory potential.  The different contributions come from the 3 terms in the potential in Eq.~\ref{eq:splitting}.  They are complex-valued in general so the magnitudes to some extent add in quadrature.  This applies in particular to  $\Delta_d$ since the random disorder gives a completely random phase to this quantity. 

We will discuss the relative contributions of $\Delta_w$, $\Delta_b$, and $\Delta_d$ in a quantitative fashion in the discussion at the end of the paper. Until then we focus on $\Delta_w$. Thus

\begin{align}
\label{eq:dwformula}
\Delta_w =& 2 \, \big| \bra{\psi_{+\mathbf{k}_0}(\mathbf{r})} V_{osc} \ket{\psi_{-\mathbf{k}_0}(\mathbf{r})} \big| \\ 
=& 2 \, \big| \sum_{\mathbf{K},\mathbf{K^{'}}} c^*_+(\mathbf{K}) c_-(\mathbf {K^{'}})
 \delta_{K_{x},K_{x^{'}}} \delta_{K_{y},K_{y^{'}}}  I_w(K_{z} - K_{z^{'}}) \big|    
\end{align}
with
\begin{equation}
\label{eq:wintegral}
I_w(K_z - K_{z}^{'})=\int_{-\infty}^{\infty}e^{iQz} V_{osc}(z)\, dz 
\end{equation}
and $Q=K_z-K_z^{'}-2k_0$.  

Eqs.~\ref{eq:dwformula} - \ref{eq:wintegral} are familiar from elementary solid-state physics, specifically from the theory of the formation of energy gaps at the surfaces of Brillouin zones.  Let the sinusoidal oscillations in $V_{osc}(z) $ be characterized by a wavevector $\pm q$ and regard $I(K_z - K_{z}^{'})$ and therefore also $\Delta_w$ and $\Delta$ as a functions of $q$.  The sum over reciprocal lattice vectors in Eq.~\ref{eq:dwformula}, together with Eq.\ref{eq:peak} means there are multiple peaks in $\Delta(q)$, one every time the condition $K_z-K_z^{'} = \pm (q \pm 2k_0)$ is satisfied.  This is the key idea for engineering the potential $V_{osc}(z)$.  

Then $I$ will peak strongly when
\begin{equation}
\label{eq:peak} 
    q = \pm Q = \pm ( K_z-K_z^{'}-2k_0), 
\end{equation}
and this in general has many solutions for $q$ since $\mathbf{K}$ and $\mathbf{K}'$ run over the reciprocal lattice.

A relative simple physical picture emerges from these equations.  We may think of the two valley minima as forming the boundaries of a one-dimensional ``Brillouin zone''. To engineer the maximum band gap, we wish to have a potential with a wavevector $ q = \pm 2k_0$.  This then corresponds to the term $K_z-K_z^{'}=0$ in the sum.  However, the Fourier transform of the cell-periodic part of the Bloch function contains all the reciprocal lattice vectors, so we can also get maxima when  $ q = \pm 2k_0$ is satisfied "modulo" a reciprocal lattice vector, which then gives the more general Eq.~\ref{eq:peak}.

Since $K_z - K_z^{'}$ is an integral multiple of $4 \pi /a$, the two shortest candidate wavevectors for the Ge oscillations from Eq.~\ref{eq:peak} are $q_1 = \pm ( 4 \pi / a - 2 k_0) $ and $q_2 = \pm 2 k_0$.  The corresponding wavelengths are $\lambda_1 = 2 \pi / q_1 = 1.80 \, \text{nm} = 13.3 \, \text{monolayers}$ and  $\lambda_2 = 2 \pi / q_2 = 0.32 \, \text{nm} = 2.36 \, \text{monolayers}$.  They correspond to what we call the long-wavelength WW and the short-wavelength WW respectively.  The former was used in Ref.~\cite{mcjunkin2021wigglewell}. Structures with wavelengths shorter than $\lambda_2$ would be difficult to fabricate, and the concept of envelope function that we use below would no longer be applicable.  These two possibilities are therefore the only ones suggested by first-order perturbation theory. Below we shall see that second-order effects give one additional candidate wavevector.      

\section{Computational Method}
\label{sec:comp}
\subsection{Introduction}
It is evident that the perturbation theory of the previous section neglects important physical effects - clearly the confinement of the electron by the electric field is not perturbative.  Thus the interplay between the localization in the z direction and the valley splitting is not properly taken into account.  Furthermore, there are three terms in the potential, and they may not all be of comparable size, so to treat them all on the same footing is not always realistic.  To remedy these defects in the simple picture, we develop a method in this section that is designed for the purpose of calculating the valley splitting in the presence of both the oscillating potential and the structure potential.  

The method has two parts: the modification of $c_{\pm}(\mathbf{K})$, the Bloch function coefficients, from their bulk values; and the calculation of the envelope functions.

\subsection{Bloch function coefficients}
\label{subsec:bloch}

A chief ingredient in the calculation of the VS in Eq.~\ref{eq:formula} is the set of the $c_{\pm}(\mathbf{K}$) defined by Eq.~\ref{eq:ck}.  We compute these coefficients using a pseudopotential method.  This is particularly appropriate for Si-Ge systems since electron energies and wavefunctions in both Si and Ge are known to be well described using just a few parameters in this formalism \cite{chelikowski1}, particularly in the energy range near the conduction band minimum. We use the local version of the method for simplicity and high throughput (which will turn out to be important).  The nonlocal version gives better bandwidths and optical matrix elements \cite{chelikowski2} and might be preferable in future work that needs higher accuracy.  Spin-orbit coupling is also neglected.  This is reasonable since we are interested only in non-magnetic properties of states near the minima of the conduction band in Si-rich materials.

The Schr\"{o}dinger equation for the Fourier components of the periodic part of the wavefunctions at the conduction band minimum wavevectors $\pm \mathbf{k}_0$ in a pure bulk system with no perturbing potential is

\begin{equation}
\label{eq:hc}
\sum_{\mathbf{K}'} H_{\mathbf{K} , \mathbf{K}'}(\mathbf{k}_0) \, c_{\pm} (\mathbf{K}')
 = \varepsilon \, c_{\pm}(\mathbf{K})
\end{equation}
The sum runs over the bcc reciprocal lattice.  In our numerical work we keep 59 terms in the sum, corresponding to the inequality $|\mathbf{K^{'}}| \leq 2 \sqrt{19} \pi/a$.  Since we are only interested in the solutions at $\mathbf{k}_0$, we will drop this argument in the remainder of this section.  $\varepsilon$ are the energies of the different bands at $\mathbf{k}_0$, each of which corresponds to an eigenvector $c({\mathbf{K}})$. The Hamiltonian matrix is 

\begin{equation}
H_{\mathbf{K} , \mathbf{K}'} = \delta_{\mathbf{K} , \mathbf{K}'} \frac{\hbar^2}{2m_e} (\mathbf{k}_0-\mathbf{K})^2 + U_{\mathbf{K} -\mathbf{K}^{'}}.
\end{equation}
$U(\mathbf r)$ is the crystal pseudopotential and  
\begin{equation}
  U_{\mathbf{K}}=\frac{1}{\nu}\int_{cell}U(\mathbf{r})e^{i\mathbf {K}\cdot\mathbf{r}}d^3r 
\end{equation}
where $\nu$ is the volume of a unit cell and the integral runs over a unit cell.

In pure Si or Ge there are two identical atoms in the unit cell and we have   
\begin{equation}
\label{eq:uc}
  U_{\mathbf{K}} = 2 \cos (\mathbf{K} \cdot \mathbf{r}_0/2) \, \frac{1}{\nu}\int_{cell}V(\mathbf {r})e^{i\mathbf K\cdot\mathbf {r}}d^3r ,
\end{equation}
where the first factor is the structure factor, $V(\mathbf{r})$ is the pseudopotential $V_{Si}$ for a single Si or $V_{Ge}$ for a Ge atom, and $\mathbf{r}_0 = (a/4) (1,1,1)$ is the separation vector of the atoms in unit cell.  We have taken the origin at the center of inversion midway between the atoms in a unit cell.  In these coordinates the $c_{\pm}(\mathbf{K})$ are real.  We have solved Eq.~\ref{eq:hc} for $c_+(\mathbf{K})$ using empirical values of the Fourier coefficients of $V(\mathbf{r})$ from Ref.~\cite{chelikowski1} and these agree with those calculated using density functional theory \cite{saraiva} on average to within 0.26\%, an accuracy that is more than enough for our purposes.  (We note that Ref.~\cite{saraiva} takes the origin at an atomic position and the resulting $c_{\pm}(\mathbf{K})$ are complex.)

To model the system with added Ge, the simplest option would be to use the standard virtual crystal approximation (VCA):
\begin{equation}
\label{eq:vch}
 H(s) = (1-x) H_{Si} + x H_{Ge}.
 \end{equation}

For a unit cell with one Si atom and one Ge atom, Eq.~\ref{eq:uc} becomes  

\begin{equation}
\label{eq:uc1}
  U_{\mathbf{K}}= 2  \cos (\mathbf{K} \cdot \mathbf{r}_0/2) \overline{V} _{\mathbf{K}} \mp i  \sin (\mathbf{K} \cdot \mathbf{r}_0/2) \delta V_{\mathbf{K}},
  \end{equation}
  where $\overline{V} _{\mathbf{K}}$ is the average of the Si and Ge pseudopotentials $V_{Si}$ and $V_{Ge}$ and $\delta V_{\mathbf{K}}$ is the difference $V_{Si} - V_{Ge}$.  The relative sign of the two terms in $U_{\mathbf{K}}$ is $\mp$ for the Si atom at $ \mp \mathbf{r}_0/2$ in the unit cell.  
  In the disordered system for each unit cell with exactly one Si atom and exactly one Ge atom these two configurations are equally probable.  
  
  The standard VCA replaces the potential of every atom with a linear combination of the potentials of a Si atom and a Ge atom.  This has the disadvantage that it artificially enforces an inversion symmetry (equal atomic potentials in the unit cell) that is not present in the real disordered system.  This turn out to be insufficient for the calculation of the VS for the long wavelength WW. To remedy this deficiency, we sample an ensemble of systems in which the positions of the Si and Ge atoms in the unit cell are random.  This is done as follows.  In a Si$_{1-x}$Ge$_{x}$ system the fraction of unit cells with exactly one Si atom and one Ge atom is $2x(1-x)$ while cells with two Si atoms have probability $(1-x)^2$ and two Ge atoms with probability $x^2$.  We treat the disordered system using an extended VCA Hamiltonian 
\begin{equation}
\label{eq:vch1}
 H(s) = (1-x)^2 H_{Si} + x^2 H_{Ge} + 2x(1-x) H_a(s).
 \end{equation}
In this equation, $H_{Si}$ and $H_{Ge}$ are the Hamiltonians for pure silicon (using only $V_{Si}$) and pure germanium (using only $V_{Ge}$), respectively. $H_a(s)$ is the alloy Hamiltonian.  In the $59 \times 59$ matrix that represents $H_a(s)$, each $s$ labels a choice of $\pm$ signs that give one realization of the disorder in the unit cell, as seen in Eq.~\ref{eq:uc1}.   Each such choice contains an equal number of plus and minus signs, and each is equally probable.  Since we can only sample a subset of these choices, we take the probability of a given $s$ be $P(s) = 1/N_0$ and for our calculations we fix $N_0$ = 300.  We then compute the density matrix 

\begin{equation}
\label{eq:rho}
\rho_{\mathbf{K},\mathbf{K}'} = \sum_{s} P(s) c^*_+(\mathbf{K},s) c_{-}(\mathbf{K}',s)
\delta_{K_{x},K_{x^{'}}} \delta_{K_{y},K_{y^{'}}} ,
  \end{equation}   
where $c_{\pm} (\mathbf{K},s)$ are the coefficients belonging to a wavefunction at the bottom of the conduction band calculated using the Hamiltonian Eq.~\ref{eq:vch1} at a fixed $s$.  $\rho_{\mathbf{K},\mathbf{K}'}$ will be a key ingredient of the computation of the VS.

\subsection{Envelope Function}
\label{subsec:envelope}

  The localization of the electron by $V_{str}(z)$ changes the wavefunctions $\psi_{+\mathbf{k}_0}(\mathbf{r})$. The widths of the wavefunctions in position and momentum space are important for computing the VS, an effect that was neglected in the derivation of Eq.~\ref{eq:integral}. The formalism we use to remedy these problems is a modification of the classic envelope method of Kohn ~\cite{kohn}.

The ordered part of the potential is 
\begin{equation}
    V(z) = V_{str}(z) + V_{osc}(z)
\end{equation}
and the total wavefunction $\Psi(\mathbf{r})$ satisfies
\begin{equation}
\label{eq:sch}
    H \Psi(\mathbf{r}) = (H_{cr} + V ) \Psi(\mathbf{r}) = E \Psi(\mathbf{r}). 
\end{equation}
In this section we neglect the disorder potential.  We will comment on this below.

We seek solutions in the form
\begin{equation}
\label{eq:psi}
	\Psi(x)=\sum_{k}A_{k} u_{k}(x)e^{ik\cdot x}
\end{equation}
The envelope function itself is
\begin{equation}
\label{eq:env}
    F(\mathbf{r}) = \sum_{\mathbf{k}} A_{\mathbf{k} } e^{-i \mathbf{k}\cdot \mathbf{r}}
\end{equation}
Substituting Eq.~\ref{eq:psi} into Eq.~\ref{eq:sch} and using Eq.~\ref{eq:crystal} we obtain the Schr\"{o}dinger equation in momentum space
\begin{equation}
\label{eq:ak}
\varepsilon_{\mathbf{k}} A_{\mathbf{k}} + \sum_{\mathbf{k}'} V_{\mathbf{k}-\mathbf{k}'} A_{\mathbf{k}} = E A_{\mathbf{k}}.
\end{equation}
$V_{\mathbf{k}-\mathbf{k}'}$ is the matrix element of the smooth potential between Bloch functions:
\begin{equation}
 V_{\mathbf{k}-\mathbf{k}'} = 
 \sum_{\mathbf{K},\mathbf{K}'} \rho_{\mathbf{K}, \mathbf{K}'} 
 \int d^3 r \, e^{i (\mathbf{K}' - \mathbf{K} + \mathbf{k}' - \mathbf{k}) \cdot \mathbf{r}} V(\mathbf{r})
\end{equation}
where Eqs.~\ref{eq:ck} and \ref{eq:rho} have been used.

So far this is quite general.  The special feature of our problem is that the wavefunction in momentum space is concentrated in the regions near $\mathbf{k} = \pm (0,0,k_0) = \pm \mathbf{k}_0$.  So we write
\begin{equation}
    F(\mathbf{r}) = F^+(\mathbf{r}) + F^-(\mathbf{r}) 
    = \sum_{\mathbf{k}\approx \mathbf{k}_0} A^+_{\mathbf{k} } e^{-i \mathbf{k}\cdot \mathbf{r}}
   + \sum_{\mathbf{k}\approx -\mathbf{k}_0} A^-_{\mathbf{k} } e^{-i \mathbf{k}\cdot \mathbf{r}}.
\end{equation}
Here $A^{\pm}_{\mathbf{k} }$ represents the function $A^{\pm}_{\mathbf{k} }$ near $\pm \mathbf{k}_0$.  More precisely, $A^{\pm}_{\mathbf{k} = \pm \mathbf{k}_0 + \mathbf{p}} =0 $, unless $|\mathbf{p}| \approx 1/Z_w << 1/a$, where $Z_w $ is the width of the envelope function in real space.

We deal for the moment only with systems that have translational invariance in the x and y directions, so $V(\mathbf {x}) = V(z)$.  Hence we may also write $F(\mathbf{r}) = F(z) $.  This excludes the possibility of treating lateral inhomogeneities such as steps in the barrier.  The current method is applicable to such problems with certain modifications, but we do not pursue this direction in this paper. 

In the presence of the oscillatory potential $ V_{\mathbf{k}-\mathbf{k}'}$ considered as a function of $\mathbf{k}$ and $\mathbf{k}'$ has 2 important regions in $\mathbf{k}$-space.

Region 1: $\mathbf{k} \approx \mathbf{k}_0$ and $\mathbf{k}' \approx -\mathbf{k}_0$.  Then
\begin{equation}
    \begin{split}
\label{eq:vc}
V^{+-}_{\mathbf{k}-\mathbf{k}'} 
& = 
 \sum_{\mathbf{K},\mathbf{K}'} \rho_{\mathbf{K},\mathbf{K}'} 
 \int d^3 r \, e^{i (\mathbf{K}' - \mathbf{K} + \mathbf{k}' - \mathbf{k}) \cdot \mathbf{r}} V(Z) \\
 & =
 \delta(k_x - k'_x)  \delta(k_y - k'_y) \\
 & \times
  \sum_{\mathbf{K},\mathbf{K}'} \rho_{\mathbf{K},\mathbf{K}'}
 \int dz e^{i (\mathbf{k}'_z - \mathbf{k}_z - 2k_0) z} V(z) 
\end{split}
\end{equation}

Region 2: $\mathbf{k} \approx -\mathbf{k}_0$ and $\mathbf{k}' \approx \mathbf{k}_0$.  
We have the simplification
\begin{equation}
V^{-+}_{\mathbf{k}-\mathbf{k}'} = ( V^{+-}_{\mathbf{k}-\mathbf{k}'} ) ^*,
\end{equation}
which enables us to write
\begin{equation}
    V_{\mathbf{k}-\mathbf{k}'}= V^{+-}_{\mathbf{k}-\mathbf{k}'}
    +V^{-+}_{\mathbf{k}-\mathbf{k}'} 
\end{equation}

 $V^{+-}_{\mathbf{k}-\mathbf{k}'}$ and $V^{-+}_{\mathbf{k}-\mathbf{k}'}$ are the parts of the potential that govern intervalley coupling and they determine the valley splitting. They depend on the density matrix $\rho_{\mathbf{K},\mathbf{K'}}$ from Eq.~\ref{eq:rho}.
 
This procedure now allows us to decompose momentum space into positive and negative $k_z$ and separate the two valleys. Eq.~\ref{eq:ak} now gives

\begin{equation}
\label{eq:akplus}
\varepsilon_{\mathbf{k}} A^+_{\mathbf{k}} 
+  \sum_{\mathbf{k}'\approx -\mathbf{k}_0} V^{+-} _{\mathbf{k}-\mathbf{k}'} A^-_{\mathbf{k}} = E A^+_{\mathbf{k}}.
\end{equation}
and
\begin{equation}
\label{eq:akminus}
\varepsilon_{\mathbf{k}} A^-_{\mathbf{k}} 
+ \sum_{\mathbf{k}'\approx \mathbf{k}_0} V^{-+} _{\mathbf{k}-\mathbf{k}'} A^+_{\mathbf{k}}  = E A^-_{\mathbf{k}}.
\end{equation}
$E$ is the total energy that includes both the barrier and the WW contributions to the valley splitting.

Finally, transforming Eqs.~\ref{eq:akplus} and \ref{eq:akminus} back to real space using Eq.~\ref{eq:env} gives a set of coupled equations for the envelope functions:
\begin{equation}
\label{eq:csw}
H_{env} \begin{pmatrix}
    F^+(z) \\
    F^-(z)
      \end{pmatrix} = 
      E
      \begin{pmatrix}
    F^+(z) \\
    F^-(z)
      \end{pmatrix}
\end{equation}
with
\begin{equation}
    \label{eq:coupled}
    H_{env} = \begin{pmatrix}
    - \frac{\hbar^2}{2m_z} \nabla^2 + V(z) & V_c(z)  \\
    (V_c(z))^* & -\frac{\hbar^2}{2m_z} \nabla^2 + V(z)
    \end{pmatrix}.
\end{equation}
Here $m_z$ is the longitudinal mass. $V_c(z)$ is the inverse Fourier transform of $V^{\pm} _{\mathbf{k}-\mathbf{k}'}$:
\begin{equation}
V_c(z) =    
\sum_{\mathbf{K},\mathbf{K}'} \rho_{\mathbf{K},\mathbf{K}'} 
V(z)
e^{i(K_z'-K_z-2k_0)z}
\delta_{K_x,K_x'}
\delta_{K_y,K_y'}.
\end{equation}
$F^{\pm}$ are the envelope functions for the $\pm k_0$ valleys. Eqs.~\ref{eq:csw} and \ref{eq:coupled} are the basic results of this section.  

The difference in the two lowest eigenvalues is the valley splitting, which now includes both barrier and WW effects.  The eigenfunction belonging to the
lowest eigenvalue determines the ground state envelope function $F(z)=F^+(z)+F^-(z)$. 

\section{Selection Rules}
\label{sec:selection}
So far the picture expected is that $\Delta_w(q)$ should have peaks of comparable sizes when $q= \pm K_z-K_z'-2k_0$ for reciprocal lattice vectors $\mathbf{K}$ and $\mathbf{K}'$ and for no other values of $q$.  It turns out, however, that this picture needs to be modified because of a selection rule that suppresses the peak at $q=3.7$nm (the long-wavelength WW). The suppression is complete if the system is considered to have the inversion symmetry in each unit cell, as would be the case in pure Si, pure Ge or the Si$_{1-x}$Ge$_x$ alloy in the standard VCA.   

This rule is derived in this section. 

As we have seen the WW part of the VS as a function of q satisfies
\begin{equation}
\label{eq:dfw}
\Delta_w(q) \propto
\big|\sum_{\mathbf{K},\mathbf{K^{'}}} c^*_+(\mathbf{K}) c_-(\mathbf {K^{'}})
 \delta_{K_{x},K_{x^{'}}} \delta_{K_{y},K_{y^{'}}} 
 \delta_{K_{z},K_{z'}+2k_0+q} \big|,  
\end{equation}
where $q$ is the wavenumber of the Ge concentration oscillation. 

We first note that for the short wavelength WW there is no selection rule.  It has $q=-2k_0$ so $K_z = K_z'$ and substituting in Eq.~\ref{eq:dwformula} yields  
\begin{equation}
\label{eq:wformula}
\Delta_w(q=-2k_0) \propto
\sum_{\mathbf{K}} |c_+(\mathbf{K})|^2 >0.   
\end{equation}
which is nonzero.  We have used the fact that $c_{-}(\mathbf{K})=c_{+}(\mathbf{K})$.

For the long wavelength WW, we define $\mathbf{G} = (4{\pi} / a) \hat{z}$ so $G=4{\pi}/a $ and we  have 
\begin{equation}
\label{eq:wformula1}
\Delta_w(G - 2k_0) \propto
\big|\sum_{\mathbf{K}} c^*_+(\mathbf{K}) c_-(\mathbf{K}+\mathbf{G}) \big|.
\end{equation}
The selection rule question boils down to the possible vanishing of the sum, which after some rearrangement is 
\begin{align}
\label{eq:wformula2}
S = \sum_{\mathbf{K}}  c_+^*(\mathbf{K}+\mathbf{G}) c_+(\mathbf{K})
\end{align}

We now demonstrate that $S=0$ for the ordered diamond structure, \textit{i.e.}, for pure Si, pure Ge, or for Si$_{1-x}$Ge$_{x}$ in the standard VCA.  

In this section it is more convenient to choose the origin at the position of an atom, which means that the $c(\mathbf{K})$ are not necessarily real. 

The symmetry group for wavevectors in the direction from $\Gamma$ to X in the Brioullin zone is $\mathbf{\Delta}$, written in bold to distinguish it from the valley splitting.  The conduction band belongs to the $\mathbf{\Delta_{1}}$ representation.  This is the identity representation for $\mathbf{\Delta}$, meaning that the lattice-periodic part of the Bloch function $u_{k}(\mathbf{r})$ is invariant under all the operations $\mathcal{U}$ of $\mathbf{\Delta}$. Hence
\begin{equation}
u_{\mathbf{k}_0}(\mathbf{r}) =  u_{\mathbf{k}_0} (\mathcal{U} \mathbf{r}),   
\end{equation}
which in turn implies that 
\begin{equation}
c(\mathcal{U}\mathbf{K}) = c(\mathbf{K})
\end{equation}
The "+" subscript will be dropped in this section for brevity, since we are only concerned with the point $ + \mathbf{k}_0$ and indeed only the conduction band.  

$\mathcal{U}$ is a product of a rotation (which may be proper or improper) and a translation.  Since there is a glide plane in Si there are  symmetries that involve a translation $\mathcal{T}$ through the vector $(a/4)(1,1,1)$ that does not belong to the fcc Bravais lattice.   We define $\mathcal{T} \mathbf{r} = \mathbf{r} + (a/4)(1,1,1)$ and let $\mathcal{C}_4$ be the rotation through $\pi/2$ about the z-axis.  The eight values of $\mathcal{U}$ are the identity $\mathcal{E}$, $C_4^2$, $\mathcal{R}$ and $\mathcal{R}'$, which are reflections in the $x=y$ and $x=-y$ plane respectively, $\mathcal{T} \times \mathcal{R} \times \mathcal{C}_4$, $\mathcal{T} \times \mathcal{R}' \times \mathcal{C}_4$, $\mathcal{T} \times \mathcal{C}_4$, and $\mathcal{T} \times \mathcal{C}_4^{-1}$. The group $\mathbf{\Delta}$ is isomorphic to $\mathbf{C_{4v}}$, which is the group of the wavevectors along the x-axis in a simple cubic lattice, but the action of the group elements on the coordinates is specific to the diamond structure.     
	
Let  $\mathcal{W}$ be a pure point operation.  Then in the $\mathbf{\Delta_1}$ representation we have the simple result that
\begin{equation}
\label{eq:uk}
    c(\mathbf{K}) = c( \mathcal{W} \mathbf{K})
\end{equation}
for these operations.  For the 4 mixed operations
$\mathcal{T} \times \mathcal{W}$ and we find
\begin{equation}
\label{eq:wk}
    c(\mathbf{K}) = \exp [ i (a/4) (\mathcal{W} \mathbf{K}) \cdot (1,1,1)]
\,    c( \mathcal{W} \mathbf{K}).
\end{equation}
These transformation properties mean that once $c(\mathbf{K}) $ is given for a certain value of $\mathbf{K} = (K_x,K_y,K_z) $ in the reciprocal lattice, then $c(\mathbf{K}')$ is determined for all other values in the orbit of $\mathbf{K}$ under the group $\mathbf{\Delta}$, which means all  $\mathbf{K}' = (K_x',K_y',K_z') $ with $K_z'=K_z$ and  $K_x'^2+K_y'^2 =K_x^2 + K_y^2$. 

An important consequence of these rules is that some of the $c(\mathbf{K})$ unexpectedly vanish. We choose any $\mathcal{U}$, set $K_x=K_y=0$, and we have that 
\begin{equation}
\label{eq:trans}
c((0,0,K_{z})) = e^{i K_z a /4 } c((0,0,K_{z})).   
\end{equation}	
For $K_z= 4 \pi /a$ this is only possible if  $c((0,0,4\pi/a)) = - c((0,0,4\pi/a)) = 0 $. 
However, if $K_z= 8 \pi /a$ the equation is an identity and we expect $c((0,0,8\pi/a)) \neq 0 $.  These patterns are evident in the results given in Ref.~\cite{saraiva}. 

We now let $\mathbf{G}=(0,0,4{\pi}/a)$ and compute the sum
\begin{equation}
S=\sum_{\mathbf{K}} c^{*}(\mathbf{K}+\mathbf{G}) c(\mathbf{K})
\end{equation}
 The orbits in $\mathbf{K}$-space consist of points with fixed $K_{z}$ and fixed $K_{x}^2 + K_{y}^2$.  They have either 1 element if the orbit is the origin, 4 elements if the $\mathbf{K}$ points are on the $K_{x}$ and $K_{y}$ axes or on the $K_{x} = K_{y}$ and $K_{x} = - K_{y}$ diagonals and 8 elements for all points not on these axes or diagonals.  Since the orbits $O$ exhaust all of $\mathbf{K}$-space, we can write
\begin{equation}
	S = \sum_{O} \sum_{\mathbf{K} \in {O}} c^*(\mathbf{K}+\mathbf{G}) c(\mathbf{K}),
\end{equation}
where the sum over $O$ runs over all orbits.  We will show that in fact
\begin{equation}
	\sum_{\mathbf{K}\in O} c^*(\mathbf{K}+\mathbf{G}) c(\mathbf{K}) = 0
\end{equation}
for all $O$, from which the selection rule $S=0$ follows.

We can classify the orbits by dividing the bcc reciprocal lattice into the $A$ sublattice $\mathbf{K} = (4\pi/a) (n_x,n_y,n_z) $ with the $n_i$ integers and the $B$ sublattice $\mathbf{K} = (4\pi/a) (n_x+1/2,n_y+1/2,n_z+1/2) $ with the $n_i$ integers. It will save writing henceforth to use only the integers $n_x,n_y$, and $n_z$ to label the $c$ coefficients so we take $a=4\pi$ and then $c(\mathbf{K}) = c(n_x,n_y,n_z)$ on the A sublattice and $c(\mathbf{K}) = c(n_x+1/2,n_y+1/2,n_z+1/2)$ on the B sublattice.

The point operations of the $\mathbf{\Delta}$ group keep $K_{z}$ fixed, so they also do not mix $A$ and $B$. Overall, we find 7 classes of orbits.  In $A$ we have $A1$ with $n_x=n_y=0$ (1 element), $A2$ with $(n_x,n_y)$ on the $n_x$ and $n_y$ axes so $(n_x,n_y)=(n_x,0)$ or $(n_x,n_y)=(0,n_y)$ (4 elements), $A3$ with $(n_x,n_y)$ on the $n_x = \pm n_y$ diagonals (4 elements), and finally $A4$ with $(n_x,n_y)$ in general position (8 elements). In $B$ the origin and the axes are  missing, and there are only 3 classes, $B1$, $B2$, and $B3$, with 4, 8, and 8 elements, respectively.  

The computation of the orbit sums is somewhat lengthy, so we give only the simplest example of the $A1$ orbit sum here and relegate the other six orbit sums to the appendix.  

The $A1$ class is of the form $\mathbf{K} = (4\pi/a) (0,0,n_z) $ and the orbit sum is

\begin{align*}
	S_{A1} & =  \sum_{\mathbf{K} \in A1} c^*(0,0,K_z+4 \pi/a) 
	\, c(0,0,K_z) \\ 
	& = 
	\sum_{n_z = -\infty}^{\infty} \, c^*(0,0,n_z+1) 
	\, c(0,0,n_z).
\end{align*}
Eq.~\ref{eq:trans} gives $c((0,0, n)) = e^{i\pi n} c((0,0, n))$
and so $c((0,0, n))=0$ if $n$ is odd.  In any term in the sum  either $n_z$ or $n_z+1$ is odd, so we find $c^*((0,0, n_z+1))) 
	\, c((0,0, n_z)) = 0$.  Every term in the sum vanishes so $S_{A1} =0$.

\section{Results}
\label{sec:res}

The envelope function is computed by discretizing the 2-component one-dimensional Schr\"{o}dinger equation, Eq.~\ref{eq:coupled} and solving it numerically. This gives a non-perturbative answer for the valley splitting.  The results can be qualitatively understood by noting that the main consequence of the calculation is to modify the integral for $\Delta_w$, by inserting of the envelope funciton in the integrand so that we have
\begin{equation}
\label{eq:wintegral1}
\Delta_w(q) \propto  I_w(K_z - K_{z}^{'})=\int_{-\infty}^{\infty}|\psi(z)|^2 \, e^{iQz} e^{-iqz} dz. 
\end{equation}   
The function $|\psi(z)|^2=|F^+(z)|^2 + |F^-(z)|^2$ has a finite spatial which then translates to peaks in $\Delta(q)$ with corresponding widths in wavenumber space.

\begin{figure}[h]
    \centering
    \subfigure[]
    {
    \includegraphics[width=0.4\textwidth]
    {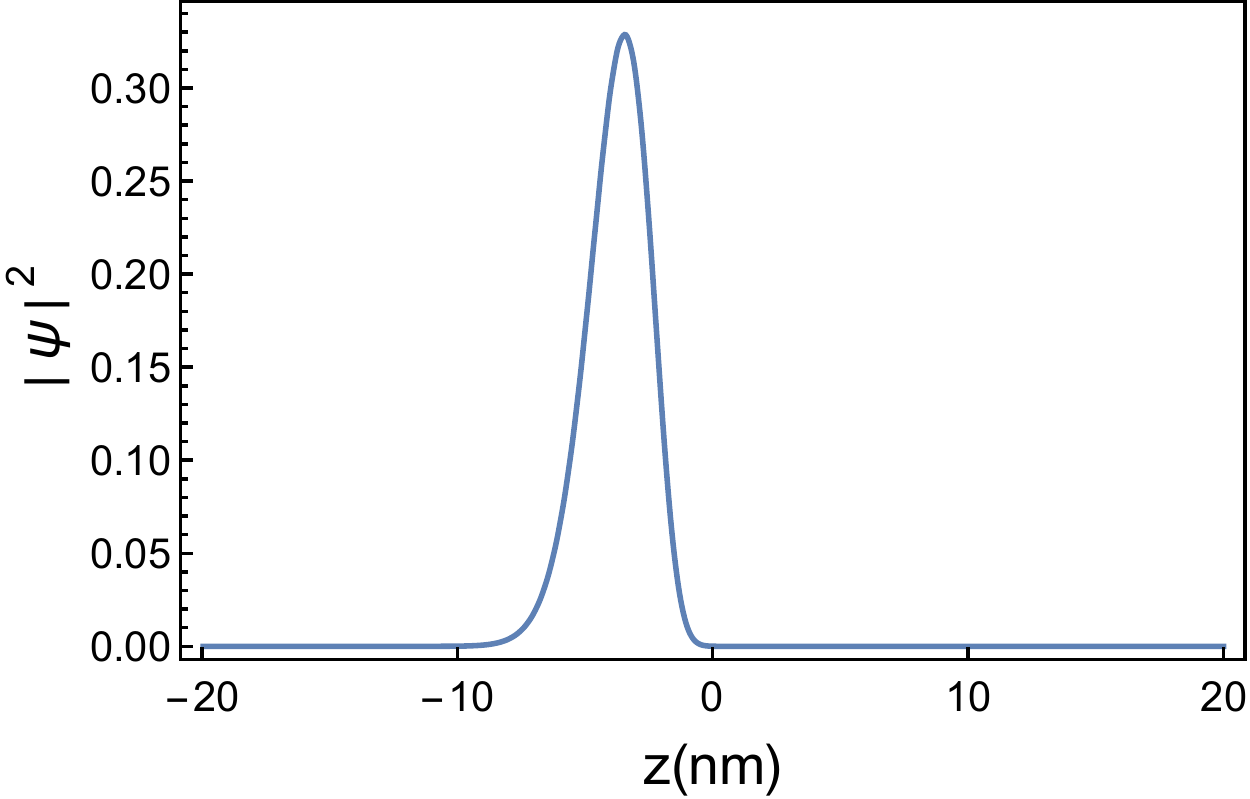}}
    \quad
    \subfigure[]{
    \includegraphics[width=0.4\textwidth]{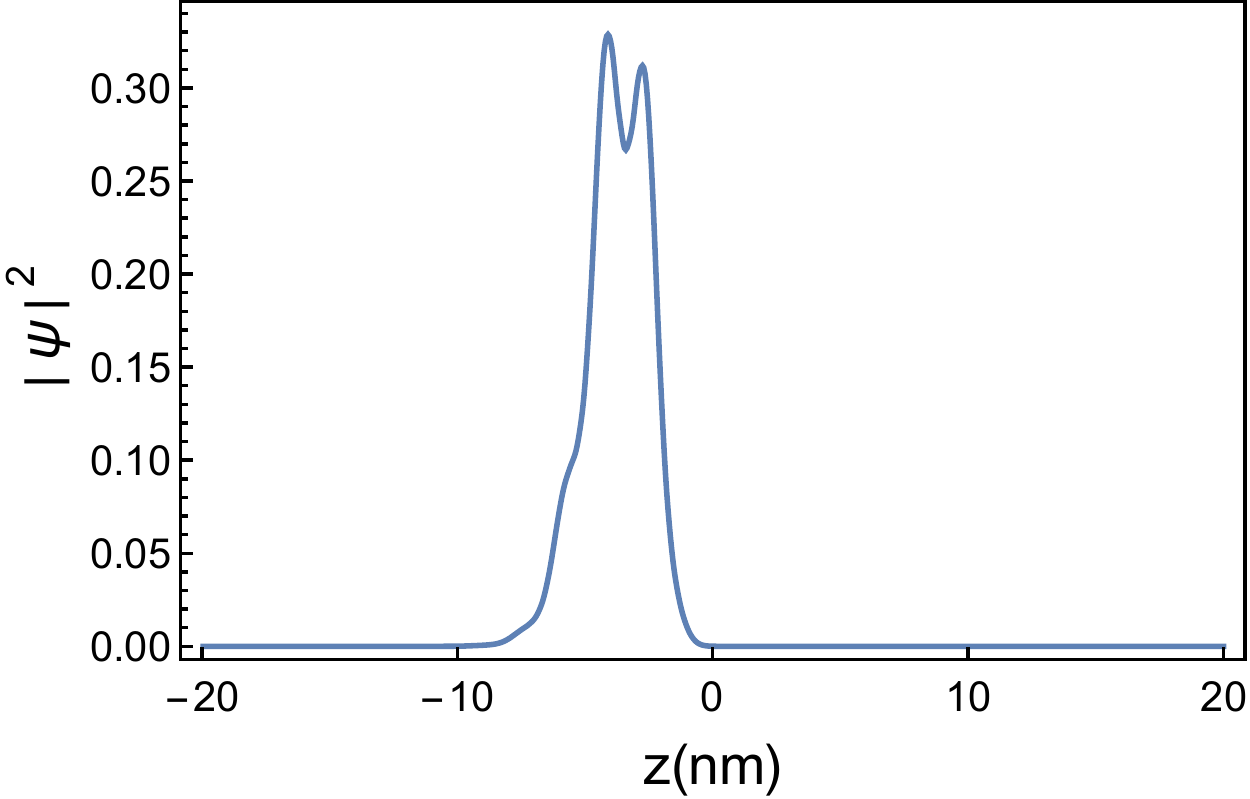}}
    \quad
    \subfigure[]{
    \includegraphics[width=0.4\textwidth]
    {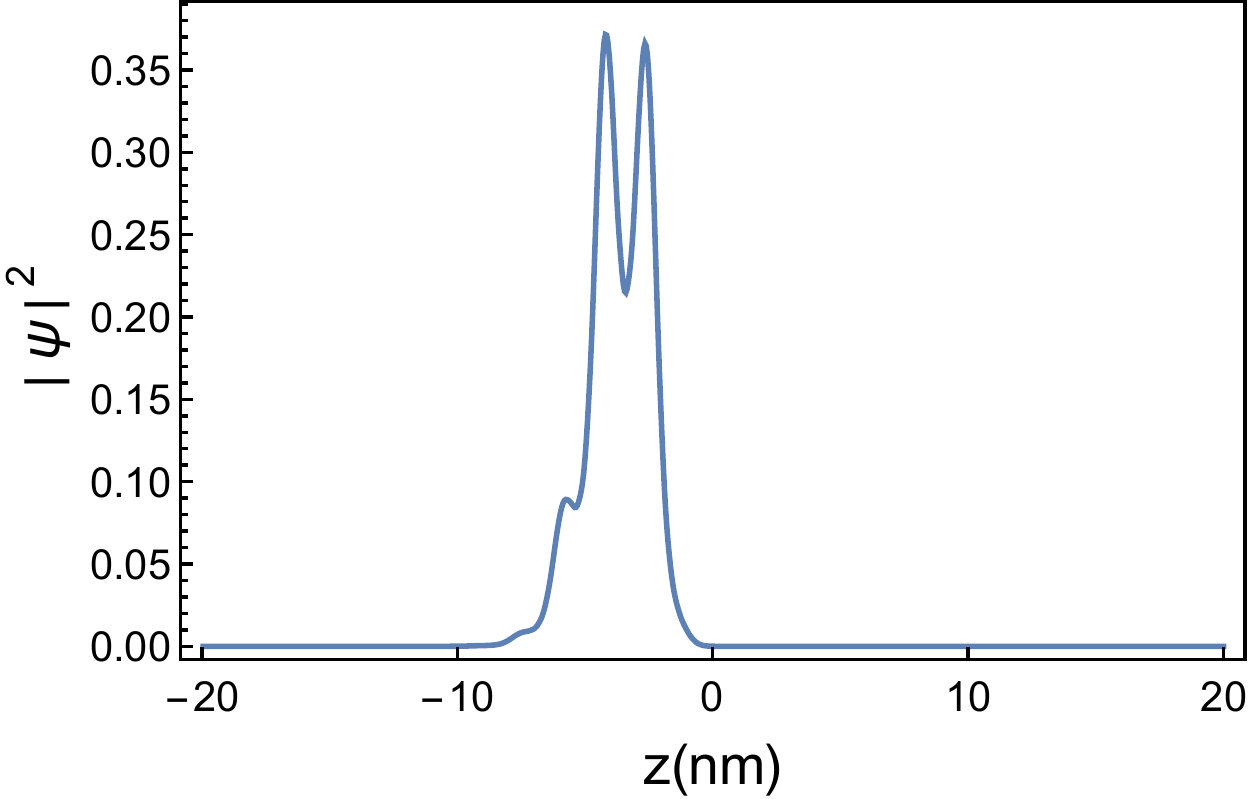}}
    \caption{Envelope functions of the ground state $|\psi(z)|^{2}=|F^{+}(z)|^{2}+|F^{-}(z)|^{2}$, plotted for $\overline{n}_{Ge}$ =0(a), 0.1(b) and 0.2(c) at $q= 3.7 \text{nm}^{-1}.$ These are the solutions of Eq.~\ref{eq:coupled}.}
    \label{fig:env}
\end{figure}

The results for the envelope function are shown in Fig.~\ref{fig:env} for $\overline{n}_{Ge}$ = 0, 0.1 and 0.2.  

The details of the envelope function depend on which device is under consideration.  However, there is a basic pattern that we expect to be universal, which is that as $\overline{n}_{Ge}$ increases, the initial single peak in $|\psi(z)|^2$ experiences increasing modulation at the period given by $q= 3.5 ~\text{nm}^{-1}$.  In this example, by the time $\overline{n}_{Ge}$ hits the rather high value 0.2, these modulations are strong enough that there are several peaks in $|\psi(z)|^2$.    

\begin{figure}[ht]
    \centering
    \subfigure[]{
    \includegraphics[width=0.4\textwidth]
    {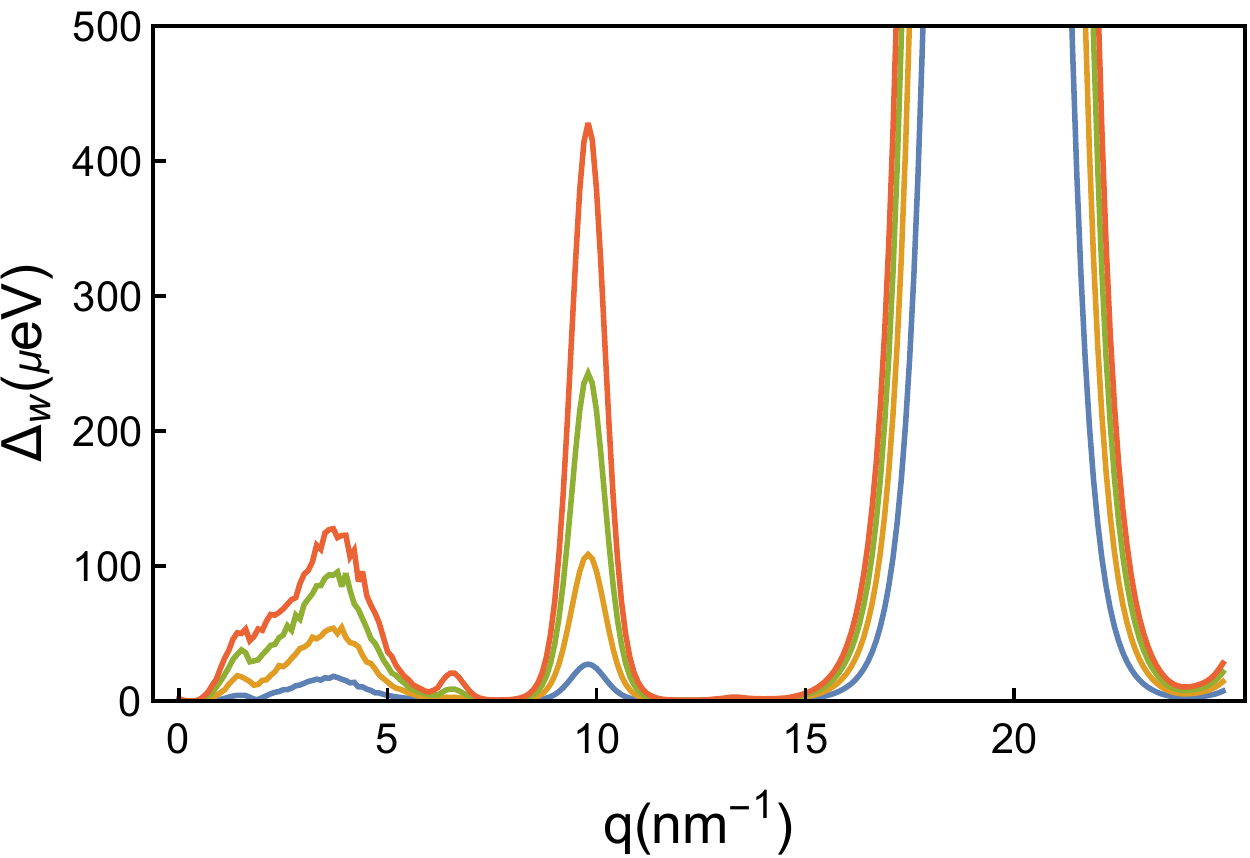}}
    \quad
    \subfigure[]{
    \includegraphics[width=0.4\textwidth]{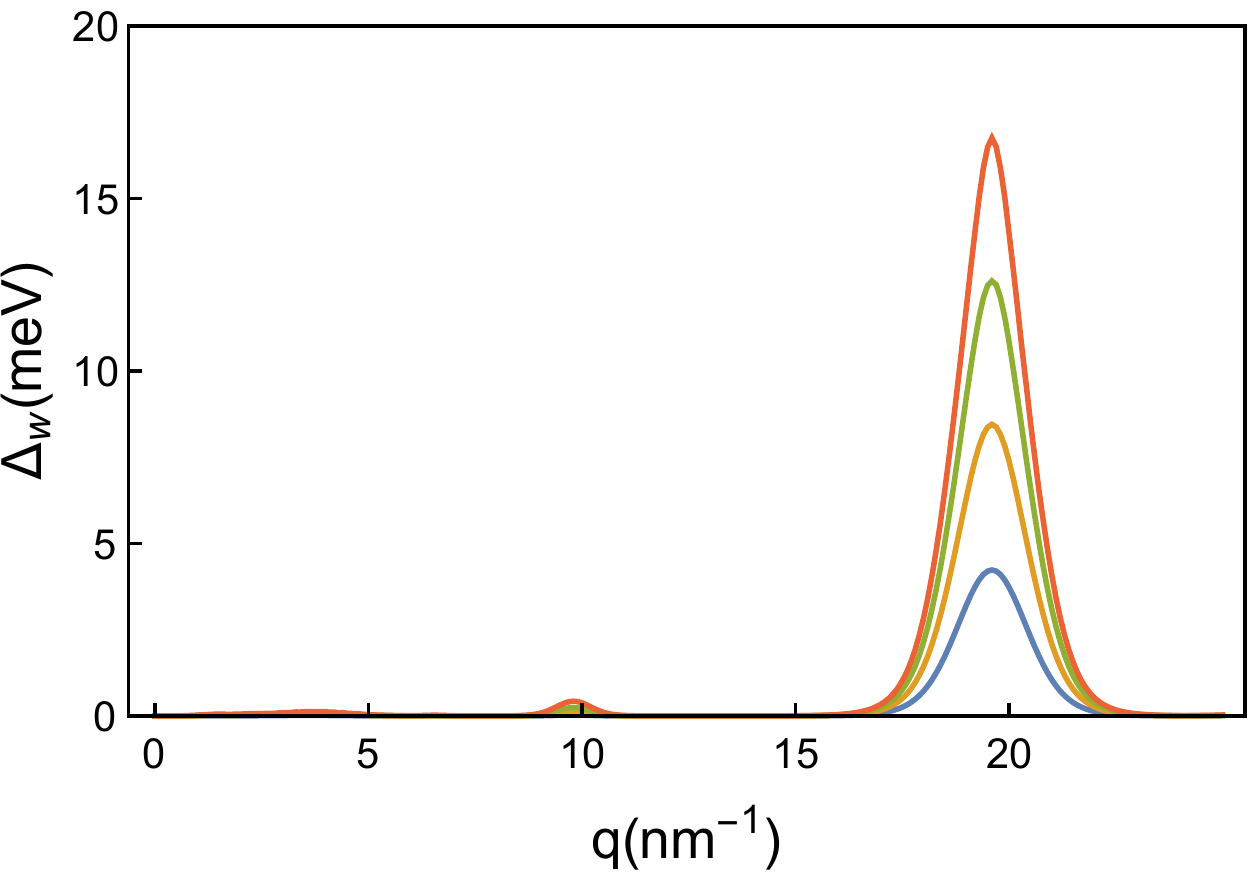}}
    \caption{The WW contribution to the valley splitting $\Delta_{w}$, plotted as a function of the wavevector of the Ge concentration $q=\frac{2\pi}{\lambda}$.  $\overline{n}_{Ge}$ = 0.05 (blue curve), 0.10 (yellow curve), 0.15 (green curve), and 0.20 (red curve). The first ($3.7\text{nm}^{-1}$) and third ($19.6\text{nm}^{-1}$) peaks correspond to the long-wavelength WW and short-wavelength WW respectively. The intermediate peak near $q=9.8\text{nm}^{-1}$ is due to a second order effect in $\overline{n}_{Ge}$. The electric field applied is $F=0.1$ V/nm.}
    \label{fig:delta}
\end{figure}

The theory we have now developed allows us to plot $\Delta_w$ versus $q = 2\pi / \lambda $, where $\lambda$ is the wavelength of the Ge concentration oscillations, versus $\overline{n}_{Ge}$, the average fractional Ge concentration in the well.  The results are shown in Fig.~\ref{fig:delta} 

The peak at small $q$ (long wavelength WW) is the one expected from the perturbative picture given above.  Its height is far smaller than that of the peak at large $q$ (short wavelength WW).  This is entirely due to the selection rule.  The fact that there is a peak at all at small $q$ is due to the fact that the disorder violates the selection rule. The noise in $\Delta_w(q)$ near $q=3.7\text{nm}^{-1}$ is due to sampling error.

In first-order perturbation theory, $|\psi(z)|^2$ has a single peak and no other structure. Hence the Fourier transform of $|\psi(z)|^2$ in Eq.~\ref{eq:wintegral1} that yields $\Delta_w(q)$ should peak only at $q= \pm K_z-K_z'-2k_0$  and the peak heights are proportional to $\overline{n}_{Ge}$.  However, the wavefunction itself develops oscillatory structure as $\overline{n}_{Ge}$ increases, as shown in Fig.~\ref{fig:env} (b) and (c).  Referring to Eq.~\ref{eq:wintegral1} and the discussion in Sec.~\ref{subsec:pert} we see that the envelope wavefunction oscillations themselves will give subsidiary peaks in $\Delta(q)$ when $q$ is one-half of the short WW value.  Thus the peak at intermediate $q$ is expected in second-order perturbation theory, when the effects of changes in wavefunctions first manifest themselves in the energy.

To complete the physical picture, we would like to verify the correctness of our contention that the physics involved in each of the three peaks shown in Fig.~\ref{fig:delta} is significantly different.  This can be done by looking at the peak height as a function of $\overline{n}_{Ge}$.  Every peak has a linear term in $\overline{n}_{Ge}$ due to the fact that the WW potential itself is proportional to $\overline{n}_{Ge}$.  The small $q$ peak is disorder-induced, implying a second factor of $(\overline{n}_{Ge})^{1/2}$, since the disorder potential results from a random walk in potential space.  Hence we expect the peak height to be proportional to $(\overline{n}_{Ge})^{3/2}$   The intermediate $q$ peak is proportional to $(\overline{n}_{Ge})^2$ because it requires wavefunction modifications that are also linear in $\overline{n}_{Ge}$, a standard perturbation theory argument. The peak at large $q$ needs no subsidiary effects for its existence so its height is linear in $\overline{n}_{Ge}$.  Of course this linearity also accounts for its much greater height.  In Fig. 4 we show the heights for each peak as a function of $\overline{n}_{Ge}$ in a log-linear plot.  The slopes of the lines confirm the overall picture very well.

\begin{figure}[h!] 
    \centering
    \includegraphics[width=0.45\textwidth]
    {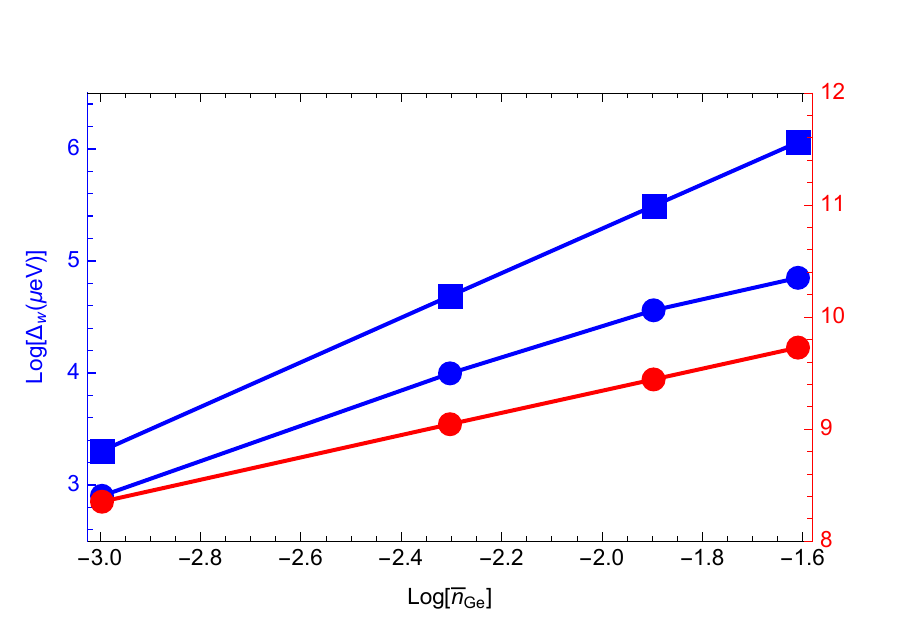}
    \caption{The logarithms of the height for each peak plotted as as a function of $\overline{n}_{Ge}$. The heights for $q = 3.7\text{nm}^{-1}$,  $q = 9.8\text{nm}^{-1}$ and $ q = 19.6\text{nm}^{-1}$ are shown by blue circles, blue squares, and red circles, respectively. The data points are connected by a guide to the eye.  Note that the points for $ q = 19.6\text{nm}^{-1}$ are referred to the vertical scale on the right.  Linear fits to the data give slopes of 1.42, 1.98, and 0.99, which agrees very well with the expected values 1.5, 2, and 1.}
    \label{fig:Maxpoint}
\end{figure}

\section{Discussion and Conclusion}
\label{sec:conclusion}
The valley splitting has three contributions: the barrier part $\Delta_b$, the disorder part $\Delta_d$, and the WW part $\Delta_w$ adding as in Eq.~\ref{eq:splitting} to give the total valley splitting $\Delta$. As we defined it, $\Delta_b$ includes \textit{all} structural effects and is the only nonzero contribution when $\overline{n}_{Ge} = 0$.  It depends on the type of device and is generally sample-dependent as well.  In practice it varies roughly from a few tens of $\mu$eV to nearly 1 meV with the high end in MOS devices. $\Delta_d$ has recently been calculated and measured in
Ref.~\cite{wuetz}.  
$\Delta_d$ ranged from 30 to 200 $\mu$eV in the experiments, with theory indicating that this could be increased to several hundred $\mu$eV by increasing the Ge concentration, particularly if the Ge atoms are inside the well.  $\Delta_d$ is of course strongly random, it being a disorder effect. Furthermore, since the positions of the Ge atoms in $V_d(\mathbf{r})$ are random, the Fourier transform of this function is flat, and any valley splitting from this source is independent of $q$.

Comparison with Fig.~\ref{fig:delta} then shows that $\Delta_w$ for the long wavelength WW has a value that is comparable to but not greater than other contributions for $ 0 < \overline{n}_{Ge} < 0.2$.  The second-harmonic peak at $q=10$nm$^{-1}$ is a little higher and it would be interesting to investigate devices designed with this wavevector.  The most important result is the large magnitude of the valley splitting of the short wavelength WW.  The results show that $\Delta_w (q = $20nm$^{-1})$ will dominate the other contributions even at Ge concentrations at the level of $\overline{n}_{Ge} << 0.05$ . It does not depend on randomness for its existence and sample dependence can be expected to be minimal. The short-wavelength WW architecture is promising for devices in which large valley splitting is important.

\begin{acknowledgements}
We thank B. Woods, M. Losert, M.A. Eriksson, M. Friesen, and S.N. Coppersmith for useful discussions, and A. Saraiva for providing important unpublished results.  This research was sponsored in part by the Army Research Office (ARO) under Grant Number W911NF-17-1-0274. The views and conclusions contained in this document are those of the authors and should not be interpreted as representing the official policies, either expressed or implied, of the Army Research Office (ARO), or the U.S. Government. The U.S. Government is authorized to reproduce and distribute reprints for Government purposes notwithstanding any copyright notation herein.
\end{acknowledgements}

\bigskip

\appendix

\section{Orbit Sums for Selection Rule}
\label{sec:orbit}
In this appendix we classify the orbits and show that each orbit sum vanishes.

The classification proceeds by dividing the bcc reciprocal lattice into the $A$ sublattice $\mathbf{K} = (4\pi/a) (n_x,n_y,n_z) $ with the $n_i$ integers and the $B$ sublattice $\mathbf{K} = (4\pi/a) (n_x+1/2,n_y+1/2,n_z+1/2) $ with the $n_i$ integers. It will save writing henceforth to use only the integers $n_x,n_y$, and $n_z$ to label the $c$ coefficients so we take $a=4\pi$ and then $c(\mathbf{K}) = c(n_x,n_y,n_z)$ on the A sublattice and $c(\mathbf{K}) = c(n_x+1/2,n_y+1/2,n_z+1/2)$ on the B sublattice.

The point operations of the $\mathbf{\Delta}$ group keep $K_{z}$ fixed, so they also do not mix $A$ and $B$. Overall, we find 7 classes of orbits. 

In $A$ we have: $A1$ with $n_x=n_y=0$, (1 element), $A2$ with $(n_x,n_y)$ on the $n_x$ and $n_y$ axes so $(n_x,n_y)=(n_x,0)$ or $(n_x,n_y)=(0,n_y)$, (4 elements), $A3$ with $(n_x,n_y)$ on the $n_x = \pm n_y$ diagonals so $(n_x,n_y)=(n_x,n_x)$ or $(n_x,n_y)=(n_x,-n_x)$ (4 elements), and finally $A4$ with $(n_x,n_y)$ in general position, (8 elements). 

In $B$ the origin and the axes are  missing, while the group operations preserve the parity of $n_x+n_y$.  There are only 3 classes: $B1$ with  $(n_x,n_y)$ on the diagonals, (4 elements), $B2$ with $|n_x| \neq |n_y|$ and $n_x+n_y$ even, (8 elements) and  $B3$ with $|n_x| \neq |n_y|$ and $n_x+n_y$ odd, (8 elements)    

Here we give the remaining orbit sums required to derive the selection rule, that is the vanishing of the sum in Eq.~\ref{eq:wformula2}. The orbit decomposition of the sum is 
\begin{equation}
    S = S_{A1}+S_{A2}+S_{A3}+S_{A4}+S_{B1}+S_{B2}+S_{B3}
\end{equation}
where
\begin{equation*}
S_{\mathcal{O}}  = \sum_{\mathbf{K} \in \mathcal{O}} c^*(\mathbf{K} + \mathbf{G}) c(\mathbf{K})
\end{equation*} 
and $\mathcal{O}$ is any of the orbits. 

\vfill

\onecolumngrid

\textbf{A Sublattice}. We have shown that $S_{A1}=0$ in the main text. Now we show that the other 6 vanish as well.

\textbf{A2.} Let $A2$ be the orbit of a vector $\mathbf{K} = (4 \pi/a) (0,n_y,n_z)$.  For a fixed $n_z$ this set has 4 elements: 
\begin{equation*}
A2 = \{(0,n_{y},n_{z}),(-n_{y},0,n_{z}),(0,-n_{y},n_{z}),(n_{y},0,n_{z})\},
\end{equation*}
where we have a adopted a notation in which $a=4 \pi$. For a fixed $n_z$ the summands in $S_{A2}$ have the form 
\begin{equation}
\label{eq:a2sum}
c^*(0,n_y,n_z+1) c(0,n_y,n_z)
+ 
c^*(-n_{y},0,n_{z}+1) c(-n_{y},0,n_{z})
+ 
c^* (0,-n_{y},n_{z}+1) c(0,-n_{y},n_{z})
+ 
c^*(n_{y},0,n_{z}+1) c(n_{y},0,n_{z}). 
\end{equation}
We now apply Eqs.~\ref{eq:uk} and \ref{eq:wk}.
If $n_{y}+n_{z}$ is even, then
\begin{equation*}
c(0,n_{y},n_{z})=c(-n_{y},0,n_{z})=c(0,-n_{y},n_{z})=c(n_{y},0,n_{z}),    
\end{equation*}
while if $n_{y}+n_{z}$ is odd, then
\begin{equation*}
c(0,n_{y},n_{z})=c(-n_{y},0,n_{z})=c(0,-n_{y},n_{z})=c(n_{y},0,n_{z})=0.   
\end{equation*}
Hence in each of the 4 terms in Eq.~\ref{eq:a2sum} one of the factors vanishes and hence $S_{A2}=0$.

\textbf{A3.} $A3$ is the orbit of a vector $\mathbf{K} = (4 \pi/a) (n_x,n_x,n_z)$ and consists of $\mathbf{K}$-vectors on the $n_x=\pm n_y$ diagonals at a fixed $n_z$. Letting $n=n_x$, these sets have 4 elements: 
\begin{equation*}
A3 = \{(n,n,n_{z}),(-n,n,n_{z}),(-n,-n,n_{z}),(n,-n,n_{z})\}.
\end{equation*}
The summands in $S_{A3}$ for a fixed $n_z$ have the form 
\begin{equation}
\label{eq:a3sum}
c^*(n,n,n_z+1) c(n,n,n_z)
+ 
c^*(-n,n,n_{z}+1) c(-n,n,n_{z})
+ 
c^* (-n,-n,n_{z}+1) c(-n,-n,n_{z})
+ 
c^*(n,-n,n_{z}+1) c(n,-n,n_{z}). 
\end{equation}
We now apply Eqs.~\ref{eq:uk} and \ref{eq:wk}.
If $n_{z}$ is even, then
\begin{equation*}
c(n,n,n_{z})=c(-n,n,n_{z})=c(-n,-n,n_{z})=c(n,-n,n_{z})   
\end{equation*}
while for $n_{z}$= odd
\begin{equation*}
c(n,n,n_{z})=c(-n,n,n_{z})=c(-n,-n,n_{z})=c(n,-n,n_{z})=0.
\end{equation*}
Hence in each of the 4 terms in Eq.~\ref{eq:a3sum} one of the factors vanishes and hence $S_{A3}=0$.

\textbf{A4.} $A4$ is the orbit of a general vector $\mathbf{K} = (4 \pi/a) (n_x,n_y,n_z)$ with $0\neq n_x \neq \pm n_y \neq 0$.  It consists of $\mathbf{K}$-vectors with fixed $K_x^2+K_y^2$ and fixed $n_z$. These sets have 8 elements: 
\begin{align*}
A4 & = \{(n_x,n_y,n_{z}),(-n_x,n_y,n_{z}),(-n_x,-n_{y},n_{z}),(n_{x},-n_y,n_{z}), \\ 
& \, \, \, \, \, \, \, \, \, (n_y,n_x,n_{z}),(-n_y,n_x,n_{z}),(-n_y,-n_{x},n_{z}),(n_{y},-n_x,n_{z})\}.  
\end{align*}
The summands in $S_{A4}$ for a fixed $n_z$ have the form 
\begin{equation}
\label{eq:a4sum}
\begin{split}
& c^*(n_x,n_y,n_z+1) c(n_x,n_y,n_z)
+ 
c^*(-n_x,n_y,n_z+1) c(-n_x,n_y,n_z) + \\
& c^* (-n_x,-n_{y},n_{z}+1) c(-n_x,-n_{y},n_{z})
+ 
c^*(n_{x},-n_y,n_{z}+1) c(n_{x},-n_y,n_{z})+ \\
& c^*(n_y,n_x,n_z+1) c(n_y,n_x,n_z)
+ 
c^*(-n_{y},n_x,n_{z}+1) c(-n_{y},n_x,n_{z})+ \\
& c^* (-n_y,-n_{x},n_{z}+1) c(-n_y,-n_{x},n_{z})
+ 
c^*(n_{y},-n_x,n_{z}+1) c(n_{y},-n_x,n_{z}). 
\end{split}
\end{equation}
For $n_z$ even and $n_x+n_y$ even we have 
\begin{align*}
c(n_{x},n_{y},n_{z}) =   c(n_{x},-n_{y},n_{z}) = 
c(-n_{x},n_{y},n_{z}) =   c(n_{y},-n_{x},n_{z}) = \\
\, \, \, \, \, \, \, c(-n_{y},n_{x},n_{z})  =  c(-n_{x},-n_{y},n_{z}) = 
c(n_{y},n_{x},n_{z}) = c(-n_{y},-n_{x},n_{z}).
\end{align*}
while for $n_z$ odd and $n_x+n_y$ even we have 
\begin{align*}
c(n_{x},n_{y},n_{z}) =   -c(n_{x},-n_{y},n_{z}) = 
-c(-n_{x},n_{y},n_{z}) =   -c(n_{y},-n_{x},n_{z}) = \\
\, \, \, \, \, \, \, -c(-n_{y},n_{x},n_{z})  =  c(-n_{x},-n_{y},n_{z}) = 
c(n_{y},n_{x},n_{z}) = c(-n_{y},-n_{x},n_{z}).
\end{align*}
Thus the 8 terms in Eq.~\ref{eq:a4sum} will completely cancel.  The sum with $n_x+n_y$ odd is similarly zero so we have $S_{A4}=0$.  

\textbf{B Sublattice.}
On the B sublattice we need to consider $c(\mathbf{K})= c[((4\pi)/a)(n_{x}+1/2,n_{y}+1/2,n_{z}+1/2)]$ which we abbreviate as $c^{\prime}(n_x,n_y,n_z)$.  

\textbf{B1.} This is the orbit of a point with $n_{x}=n_{y}$. The orbit has 4 elements: 
\begin{equation*}
  B1 = \{ (n_{x},n_{x},n_{z}), (-n_{x},n_{x},n_{z}),(n_{x},-n_{x},n_{z}),(-n_{x},-n_{x},n_{z} \} .
\end{equation*}
The summands in $S_{B1}$ have the form
\begin{equation}
\label{eq:b1}
\begin{split}
 & c'^*(n_{x},n_{x},n_{z}+1)c'(n_{x},n_{x},n_{z})+c'^*(-n_{x},-n_{x},n_{z}+1)c'(-n_{x},-n_{x},n_{z}) + \\
&c'^*(n_{x},-n_{x},n_{z}+1)c'(n_{x},-n_{x},n_{z})+c'^*(-n_{x},n_{x},n_{z}+1)c'(-n_{x},n_{x},n_{z})
\end{split}
\end{equation}

If $n_{z}$ is even we have
\begin{equation*}
c'(n_{x},n_{x},n_{z}) = c'(-n_{x},-n_{x},n_{z})
= -ic'(n_{x},-n_{x},n_{z})=-ic'(-n_{x},n_{x},n_{z})
\end{equation*}
while if $n_z$ is odd then
\begin{equation*}
c'(n_{x},n_{x},n_{z}) = c'(-n_{x},-n_{x},n_{z})
= -ic'(n_{x},-n_{x},n_{z})=-ic'(-n_{x},n_{x},n_{z}).   
\end{equation*}
Substitution in Eq.~\ref{eq:b1} then shows that the orbit sum is zero.

\textbf{B2.} This is the orbit of a point with $0 \neq n_{x} \neq \pm n_{y} \neq 0$ and $n_x+n_y$ even. The orbit has 8 elements: 
\begin{align*}
  B2 = & \{ (n_{x},n_{y},n_{z}), (n_{x},-n_{y},n_{z}),
	 (-n_{x},n_{y},n_{z}),
	 (n_{y},-n_{x},n_{z}), \\
	  & (-n_{y},n_{x},n_{z}), 
	   (-n_{x},-n_{y},n_{z}),
	  (n_{y},n_{x},n_{z}),
	  (-n_{y},-n_{x},n_{z}) \}.
\end{align*}
The summands in $S_{B2}$ have the form
\begin{equation}
\begin{split}
\label{eq:b2sum}
& c'^*(n_{x},n_{y},n_{z}+1)c'(n_{x},n_{y},n_{z})
 +
c'^*(n_{x},-n_{y},n_{z}+1)c'(n_{x},-n_{y},n_{z}) + \\
& c'^*(-n_{x},n_{y},n_{z}+1)c'(-n_{x},n_{y},n_{z})
+
c'^*(n_{y},-n_{x},n_{z}+1)c'(n_{y},-n_{x},n_{z})+ \\
& c'^*(-n_{y},n_{x},n_{z}+1)c'(-n_{y},n_{x},n_{z})
+
c'^*(-n_{x},-n_{y},n_{z}+1)c'(-n_{x},-n_{y},n_{z}) + \\
& c'^*(n_{y},n_{x},n_{z}+1)c'(n_{y},n_{x},n_{z})
+
c'^*(-n_{y},-n_{x},n_{z}+1)c'(-n_{y},-n_{x},n_{z})
\end{split}
\end{equation}

If $n_{z}$ is even we have
\begin{align*}
&	c'(n_{x},n_{y},n_{z}) =  c'(n_{x},-n_{y},n_{z})
	 =  c'(-n_{x},n_{y},n_{z})
	 =  c'(n_{y},-n_{x},n_{z}) \\
&	 =  c'(-n_{y},n_{x},n_{z})
	 = c'(-n_{x},-n_{y},n_{z})
	 = c'(n_{y},n_{x},n_{z})
	 = c'(-n_{y},-n_{x},n_{z}).
\end{align*}

while if $n_z$ is odd then
\begin{align*}
&	c'(n_{x},n_{y},n_{z}) = - c'(n_{x},-n_{y},n_{z})
	 =  -c'(-n_{x},n_{y},n_{z})
	 =  -c'(n_{y},-n_{x},n_{z}) \\
&	 =  -c'(-n_{y},n_{x},n_{z})
	 = c'(-n_{x},-n_{y},n_{z})
	 = c'(n_{y},n_{x},n_{z})
	 = c'(-n_{y},-n_{x},n_{z}).
\end{align*}
Substitution in Eq.~\ref{eq:b2sum} then shows that $S_{B2}=0$.

\textbf{B3.} This is the orbit of a point with $0 \neq n_{x} \neq \pm n_{y} \neq 0$ and $n_x+n_y$ odd. The orbit has 8 elements.  The elements and the typical term in $S_{B3}$ are the same as for $B2$ but the transformation properties are different. 

If $n_{z}$ is even we have
\begin{align*}
& c'(n_{x},n_{y},n_{z}) = - c'(n_{x},-n_{y},n_{z})
	 = - c'(-n_{x},n_{y},n_{z})
	 = - c'(n_{y},-n_{x},n_{z}) \\
&	 = -c'(-n_{y},n_{x},n_{z}) 
	 = c'(-n_{x},-n_{y},n_{z})
	 = c'(n_{y},n_{x},n_{z})
	 = c'(-n_{y},-n_{x},n_{z}).
\end{align*}
while if $n_z$ is odd then
\begin{align*}
&	c'(n_{x},n_{y},n_{z}) = c'(n_{x},-n_{y},n_{z})
	 = c'(-n_{x},n_{y},n_{z})
	 = c'(n_{y},-n_{x},n_{z}) \\
&	 = c'(-n_{y},n_{x},n_{z})
	 = c'(-n_{x},-n_{y},n_{z})
	 = c'(n_{y},n_{x},n_{z})
	 = c'(-n_{y},-n_{x},n_{z}).
	 \end{align*}
Also in this case, substitution of these relations into Eq.~\ref{eq:b2sum} shows that $S_{B3}=0$.

This completes the proof.

\bibliography{main}

\providecommand{\noopsort}[1]{}\providecommand{\singleletter}[1]{#1}%
\begin{thebibliography}{22}%
\makeatletter
\providecommand \@ifxundefined [1]{%
 \@ifx{#1\undefined}
}%
\providecommand \@ifnum [1]{%
 \ifnum #1\expandafter \@firstoftwo
 \else \expandafter \@secondoftwo
 \fi
}%
\providecommand \@ifx [1]{%
 \ifx #1\expandafter \@firstoftwo
 \else \expandafter \@secondoftwo
 \fi
}%
\providecommand \natexlab [1]{#1}%
\providecommand \enquote  [1]{``#1''}%
\providecommand \bibnamefont  [1]{#1}%
\providecommand \bibfnamefont [1]{#1}%
\providecommand \citenamefont [1]{#1}%
\providecommand \href@noop [0]{\@secondoftwo}%
\providecommand \href [0]{\begingroup \@sanitize@url \@href}%
\providecommand \@href[1]{\@@startlink{#1}\@@href}%
\providecommand \@@href[1]{\endgroup#1\@@endlink}%
\providecommand \@sanitize@url [0]{\catcode `\\12\catcode `\$12\catcode
  `\&12\catcode `\#12\catcode `\^12\catcode `\_12\catcode `\%12\relax}%
\providecommand \@@startlink[1]{}%
\providecommand \@@endlink[0]{}%
\providecommand \url  [0]{\begingroup\@sanitize@url \@url }%
\providecommand \@url [1]{\endgroup\@href {#1}{\urlprefix }}%
\providecommand \urlprefix  [0]{URL }%
\providecommand \Eprint [0]{\href }%
\providecommand \doibase [0]{https://doi.org/}%
\providecommand \selectlanguage [0]{\@gobble}%
\providecommand \bibinfo  [0]{\@secondoftwo}%
\providecommand \bibfield  [0]{\@secondoftwo}%
\providecommand \translation [1]{[#1]}%
\providecommand \BibitemOpen [0]{}%
\providecommand \bibitemStop [0]{}%
\providecommand \bibitemNoStop [0]{.\EOS\space}%
\providecommand \EOS [0]{\spacefactor3000\relax}%
\providecommand \BibitemShut  [1]{\csname bibitem#1\endcsname}%
\let\auto@bib@innerbib\@empty
\bibitem [{\citenamefont {Zwanenburg}\ \emph {et~al.}(2013)\citenamefont
  {Zwanenburg}, \citenamefont {Dzurak}, \citenamefont {Morello}, \citenamefont
  {Simmons}, \citenamefont {Hollenberg}, \citenamefont {Klimeck}, \citenamefont
  {Rogge}, \citenamefont {Coppersmith},\ and\ \citenamefont
  {Eriksson}}]{Zwanenburg:2013p961}%
  \BibitemOpen
  \bibfield  {author} {\bibinfo {author} {\bibfnamefont {F.~A.}\ \bibnamefont
  {Zwanenburg}}, \bibinfo {author} {\bibfnamefont {A.~S.}\ \bibnamefont
  {Dzurak}}, \bibinfo {author} {\bibfnamefont {A.}~\bibnamefont {Morello}},
  \bibinfo {author} {\bibfnamefont {M.~Y.}\ \bibnamefont {Simmons}}, \bibinfo
  {author} {\bibfnamefont {L.~C.~L.}\ \bibnamefont {Hollenberg}}, \bibinfo
  {author} {\bibfnamefont {G.}~\bibnamefont {Klimeck}}, \bibinfo {author}
  {\bibfnamefont {S.}~\bibnamefont {Rogge}}, \bibinfo {author} {\bibfnamefont
  {S.~N.}\ \bibnamefont {Coppersmith}},\ and\ \bibinfo {author} {\bibfnamefont
  {M.~A.}\ \bibnamefont {Eriksson}},\ }\bibfield  {title} {\bibinfo {title}
  {Silicon quantum electronics},\ }\href@noop {} {\bibfield  {journal}
  {\bibinfo  {journal} {Rev. Mod. Phys.}\ }\textbf {\bibinfo {volume} {85}},\
  \bibinfo {pages} {961} (\bibinfo {year} {2013})}\BibitemShut {NoStop}%
\bibitem [{\citenamefont {Weitz}\ \emph {et~al.}(1996)\citenamefont {Weitz},
  \citenamefont {Haug}, \citenamefont {von Klitzing},\ and\ \citenamefont
  {Sch\"{a}ffler}}]{Weitz:1996p542}%
  \BibitemOpen
  \bibfield  {author} {\bibinfo {author} {\bibfnamefont {P.}~\bibnamefont
  {Weitz}}, \bibinfo {author} {\bibfnamefont {R.}~\bibnamefont {Haug}},
  \bibinfo {author} {\bibfnamefont {K.}~\bibnamefont {von Klitzing}},\ and\
  \bibinfo {author} {\bibfnamefont {F.}~\bibnamefont {Sch\"{a}ffler}},\
  }\bibfield  {title} {\bibinfo {title} {Tilted magnetic field studies of spin-
  and valley-splittings in {S}i/{S}i$_{1-x}${G}e$_x$ heterostructures},\ }\href
  {https://doi.org/10.1016/0039-6028(96)00465-7} {\bibfield  {journal}
  {\bibinfo  {journal} {Surface Science}\ }\textbf {\bibinfo {volume}
  {361-362}},\ \bibinfo {pages} {542} (\bibinfo {year} {1996})}\BibitemShut
  {NoStop}%
\bibitem [{\citenamefont {Koester}\ \emph {et~al.}(1996)\citenamefont
  {Koester}, \citenamefont {Ismail}, \citenamefont {Lee},\ and\ \citenamefont
  {Chu}}]{Koester:1996p1400}%
  \BibitemOpen
  \bibfield  {author} {\bibinfo {author} {\bibfnamefont {S.~J.}\ \bibnamefont
  {Koester}}, \bibinfo {author} {\bibfnamefont {K.}~\bibnamefont {Ismail}},
  \bibinfo {author} {\bibfnamefont {K.~Y.}\ \bibnamefont {Lee}},\ and\ \bibinfo
  {author} {\bibfnamefont {J.~O.}\ \bibnamefont {Chu}},\ }\bibfield  {title}
  {\bibinfo {title} {Weak localization in back-gated
  {S}i/{S}i$_{0.7}${G}e$_{0.3}$ quantum-well wires fabricated by reactive ion
  etching},\ }\href {https://doi.org/10.1103/PhysRevB.54.10604} {\bibfield
  {journal} {\bibinfo  {journal} {Phys Rev B}\ }\textbf {\bibinfo {volume}
  {54}},\ \bibinfo {pages} {10604} (\bibinfo {year} {1996})}\BibitemShut
  {NoStop}%
\bibitem [{\citenamefont {Lai}\ \emph {et~al.}(2006)\citenamefont {Lai},
  \citenamefont {Lu}, \citenamefont {Pan}, \citenamefont {Tsui}, \citenamefont
  {Lyon}, \citenamefont {Liu}, \citenamefont {Xie}, \citenamefont
  {M\"{u}hlberger},\ and\ \citenamefont {Sch\"{a}ffler}}]{Lai:2006p161301}%
  \BibitemOpen
  \bibfield  {author} {\bibinfo {author} {\bibfnamefont {K.}~\bibnamefont
  {Lai}}, \bibinfo {author} {\bibfnamefont {T.~M.}\ \bibnamefont {Lu}},
  \bibinfo {author} {\bibfnamefont {W.}~\bibnamefont {Pan}}, \bibinfo {author}
  {\bibfnamefont {D.~C.}\ \bibnamefont {Tsui}}, \bibinfo {author}
  {\bibfnamefont {S.}~\bibnamefont {Lyon}}, \bibinfo {author} {\bibfnamefont
  {J.}~\bibnamefont {Liu}}, \bibinfo {author} {\bibfnamefont {Y.~H.}\
  \bibnamefont {Xie}}, \bibinfo {author} {\bibfnamefont {M.}~\bibnamefont
  {M\"{u}hlberger}},\ and\ \bibinfo {author} {\bibfnamefont {F.}~\bibnamefont
  {Sch\"{a}ffler}},\ }\bibfield  {title} {\bibinfo {title} {Valley splitting of
  {S}i/{S}i$_{1-x}${G}e$_x$ heterostructures in tilted magnetic fields},\
  }\href {https://doi.org/10.1103/PhysRevB.73.161301} {\bibfield  {journal}
  {\bibinfo  {journal} {Phys Rev B}\ }\textbf {\bibinfo {volume} {73}},\
  \bibinfo {pages} {161301(R)} (\bibinfo {year} {2006})}\BibitemShut {NoStop}%
\bibitem [{\citenamefont {Goswami}\ \emph
  {et~al.}(2007{\natexlab{a}})\citenamefont {Goswami}, \citenamefont {Slinker},
  \citenamefont {Friesen}, \citenamefont {McGuire}, \citenamefont {Truitt},
  \citenamefont {Tahan}, \citenamefont {Klein}, \citenamefont {Chu},
  \citenamefont {Mooney}, \citenamefont {van~der Weide}, \citenamefont {Joynt},
  \citenamefont {Coppersmith},\ and\ \citenamefont
  {Eriksson}}]{Goswami:2007p41}%
  \BibitemOpen
  \bibfield  {author} {\bibinfo {author} {\bibfnamefont {S.}~\bibnamefont
  {Goswami}}, \bibinfo {author} {\bibfnamefont {K.~A.}\ \bibnamefont
  {Slinker}}, \bibinfo {author} {\bibfnamefont {M.}~\bibnamefont {Friesen}},
  \bibinfo {author} {\bibfnamefont {L.~M.}\ \bibnamefont {McGuire}}, \bibinfo
  {author} {\bibfnamefont {J.~L.}\ \bibnamefont {Truitt}}, \bibinfo {author}
  {\bibfnamefont {C.}~\bibnamefont {Tahan}}, \bibinfo {author} {\bibfnamefont
  {L.~J.}\ \bibnamefont {Klein}}, \bibinfo {author} {\bibfnamefont {J.~O.}\
  \bibnamefont {Chu}}, \bibinfo {author} {\bibfnamefont {P.~M.}\ \bibnamefont
  {Mooney}}, \bibinfo {author} {\bibfnamefont {D.~W.}\ \bibnamefont {van~der
  Weide}}, \bibinfo {author} {\bibfnamefont {R.}~\bibnamefont {Joynt}},
  \bibinfo {author} {\bibfnamefont {S.~N.}\ \bibnamefont {Coppersmith}},\ and\
  \bibinfo {author} {\bibfnamefont {M.~A.}\ \bibnamefont {Eriksson}},\
  }\bibfield  {title} {\bibinfo {title} {Controllable valley splitting in
  silicon quantum devices},\ }\href {https://doi.org/10.1038/nphys475}
  {\bibfield  {journal} {\bibinfo  {journal} {Nat. Phys.}\ }\textbf {\bibinfo
  {volume} {3}},\ \bibinfo {pages} {41} (\bibinfo {year}
  {2007}{\natexlab{a}})}\BibitemShut {NoStop}%
\bibitem [{\citenamefont {Mi}\ \emph {et~al.}(2015)\citenamefont {Mi},
  \citenamefont {Hazard}, \citenamefont {Payette}, \citenamefont {Wang},
  \citenamefont {Zajac}, \citenamefont {Cady},\ and\ \citenamefont
  {Petta}}]{Mi:2015p035304}%
  \BibitemOpen
  \bibfield  {author} {\bibinfo {author} {\bibfnamefont {X.}~\bibnamefont
  {Mi}}, \bibinfo {author} {\bibfnamefont {T.~M.}\ \bibnamefont {Hazard}},
  \bibinfo {author} {\bibfnamefont {C.}~\bibnamefont {Payette}}, \bibinfo
  {author} {\bibfnamefont {K.}~\bibnamefont {Wang}}, \bibinfo {author}
  {\bibfnamefont {D.~M.}\ \bibnamefont {Zajac}}, \bibinfo {author}
  {\bibfnamefont {J.~V.}\ \bibnamefont {Cady}},\ and\ \bibinfo {author}
  {\bibfnamefont {J.~R.}\ \bibnamefont {Petta}},\ }\bibfield  {title} {\bibinfo
  {title} {Magnetotransport studies of mobility limiting mechanisms in undoped
  {S}i/{S}i{G}e heterostructures},\ }\href
  {https://doi.org/10.1103/PhysRevB.92.035304} {\bibfield  {journal} {\bibinfo
  {journal} {Phys. Rev. B}\ }\textbf {\bibinfo {volume} {92}},\ \bibinfo
  {pages} {035304} (\bibinfo {year} {2015})}\BibitemShut {NoStop}%
\bibitem [{\citenamefont {Mi}\ \emph {et~al.}(2017)\citenamefont {Mi},
  \citenamefont {P\'eterfalvi}, \citenamefont {Burkard},\ and\ \citenamefont
  {Petta}}]{MiPRL2017}%
  \BibitemOpen
  \bibfield  {author} {\bibinfo {author} {\bibfnamefont {X.}~\bibnamefont
  {Mi}}, \bibinfo {author} {\bibfnamefont {C.~G.}\ \bibnamefont
  {P\'eterfalvi}}, \bibinfo {author} {\bibfnamefont {G.}~\bibnamefont
  {Burkard}},\ and\ \bibinfo {author} {\bibfnamefont {J.~R.}\ \bibnamefont
  {Petta}},\ }\bibfield  {title} {\bibinfo {title} {High-resolution valley
  spectroscopy of si quantum dots},\ }\href
  {https://doi.org/10.1103/PhysRevLett.119.176803} {\bibfield  {journal}
  {\bibinfo  {journal} {Phys. Rev. Lett.}\ }\textbf {\bibinfo {volume} {119}},\
  \bibinfo {pages} {176803} (\bibinfo {year} {2017})}\BibitemShut {NoStop}%
\bibitem [{\citenamefont {Neyens}\ \emph {et~al.}(2018)\citenamefont {Neyens},
  \citenamefont {Foote}, \citenamefont {Thorgrimsson}, \citenamefont {Knapp},
  \citenamefont {McJunkin}, \citenamefont {Vandersypen}, \citenamefont {Amin},
  \citenamefont {Thomas}, \citenamefont {Clarke}, \citenamefont {Savage},
  \citenamefont {Lagally}, \citenamefont {Friesen}, \citenamefont
  {Coppersmith},\ and\ \citenamefont {Eriksson}}]{Neyens:2018p243107}%
  \BibitemOpen
  \bibfield  {author} {\bibinfo {author} {\bibfnamefont {S.~F.}\ \bibnamefont
  {Neyens}}, \bibinfo {author} {\bibfnamefont {R.~H.}\ \bibnamefont {Foote}},
  \bibinfo {author} {\bibfnamefont {B.}~\bibnamefont {Thorgrimsson}}, \bibinfo
  {author} {\bibfnamefont {T.~J.}\ \bibnamefont {Knapp}}, \bibinfo {author}
  {\bibfnamefont {T.}~\bibnamefont {McJunkin}}, \bibinfo {author}
  {\bibfnamefont {L.~M.~K.}\ \bibnamefont {Vandersypen}}, \bibinfo {author}
  {\bibfnamefont {P.}~\bibnamefont {Amin}}, \bibinfo {author} {\bibfnamefont
  {N.~K.}\ \bibnamefont {Thomas}}, \bibinfo {author} {\bibfnamefont {J.~S.}\
  \bibnamefont {Clarke}}, \bibinfo {author} {\bibfnamefont {D.~E.}\
  \bibnamefont {Savage}}, \bibinfo {author} {\bibfnamefont {M.~G.}\
  \bibnamefont {Lagally}}, \bibinfo {author} {\bibfnamefont {M.}~\bibnamefont
  {Friesen}}, \bibinfo {author} {\bibfnamefont {S.~N.}\ \bibnamefont
  {Coppersmith}},\ and\ \bibinfo {author} {\bibfnamefont {M.~A.}\ \bibnamefont
  {Eriksson}},\ }\bibfield  {title} {\bibinfo {title} {The critical role of
  substrate disorder in valley splitting in {S}i quantum wells},\ }\href@noop
  {} {\bibfield  {journal} {\bibinfo  {journal} {Appl. Phys. Lett.}\ }\textbf
  {\bibinfo {volume} {112}},\ \bibinfo {pages} {243107} (\bibinfo {year}
  {2018})}\BibitemShut {NoStop}%
\bibitem [{\citenamefont {Yang}\ \emph {et~al.}(2013)\citenamefont {Yang},
  \citenamefont {Rossi}, \citenamefont {Ruskov}, \citenamefont {Lai},
  \citenamefont {Mohiyaddin}, \citenamefont {Lee}, \citenamefont {Tahan},
  \citenamefont {Klimeck}, \citenamefont {Morello},\ and\ \citenamefont
  {Dzurak}}]{Yang:2013p3069}%
  \BibitemOpen
  \bibfield  {author} {\bibinfo {author} {\bibfnamefont {C.~H.}\ \bibnamefont
  {Yang}}, \bibinfo {author} {\bibfnamefont {A.}~\bibnamefont {Rossi}},
  \bibinfo {author} {\bibfnamefont {R.}~\bibnamefont {Ruskov}}, \bibinfo
  {author} {\bibfnamefont {N.~S.}\ \bibnamefont {Lai}}, \bibinfo {author}
  {\bibfnamefont {F.~A.}\ \bibnamefont {Mohiyaddin}}, \bibinfo {author}
  {\bibfnamefont {S.}~\bibnamefont {Lee}}, \bibinfo {author} {\bibfnamefont
  {C.}~\bibnamefont {Tahan}}, \bibinfo {author} {\bibfnamefont
  {G.}~\bibnamefont {Klimeck}}, \bibinfo {author} {\bibfnamefont
  {A.}~\bibnamefont {Morello}},\ and\ \bibinfo {author} {\bibfnamefont {A.~S.}\
  \bibnamefont {Dzurak}},\ }\bibfield  {title} {\bibinfo {title} {Spin-valley
  lifetimes in a silicon quantum dot with tunable valley splitting},\ }\href
  {https://doi.org/10.1038/ncomms3069} {\bibfield  {journal} {\bibinfo
  {journal} {Nature Communications}\ }\textbf {\bibinfo {volume} {4}},\
  \bibinfo {pages} {2069} (\bibinfo {year} {2013})}\BibitemShut {NoStop}%
\bibitem [{\citenamefont {Gamble}\ \emph {et~al.}(2016)\citenamefont {Gamble},
  \citenamefont {Harvey-Collard}, \citenamefont {Jacobson}, \citenamefont
  {Baczewski}, \citenamefont {Nielsen}, \citenamefont {Maurer}, \citenamefont
  {Monta{\~n}o}, \citenamefont {Rudolph}, \citenamefont {Carroll},
  \citenamefont {Yang}, \citenamefont {Rossi}, \citenamefont {Dzurak},\ and\
  \citenamefont {Muller}}]{Gamble:2016p253101}%
  \BibitemOpen
  \bibfield  {author} {\bibinfo {author} {\bibfnamefont {J.~K.}\ \bibnamefont
  {Gamble}}, \bibinfo {author} {\bibfnamefont {P.}~\bibnamefont
  {Harvey-Collard}}, \bibinfo {author} {\bibfnamefont {N.~T.}\ \bibnamefont
  {Jacobson}}, \bibinfo {author} {\bibfnamefont {A.~D.}\ \bibnamefont
  {Baczewski}}, \bibinfo {author} {\bibfnamefont {E.}~\bibnamefont {Nielsen}},
  \bibinfo {author} {\bibfnamefont {L.}~\bibnamefont {Maurer}}, \bibinfo
  {author} {\bibfnamefont {I.}~\bibnamefont {Monta{\~n}o}}, \bibinfo {author}
  {\bibfnamefont {M.}~\bibnamefont {Rudolph}}, \bibinfo {author} {\bibfnamefont
  {M.}~\bibnamefont {Carroll}}, \bibinfo {author} {\bibfnamefont
  {C.}~\bibnamefont {Yang}}, \bibinfo {author} {\bibfnamefont {A.}~\bibnamefont
  {Rossi}}, \bibinfo {author} {\bibfnamefont {A.}~\bibnamefont {Dzurak}},\ and\
  \bibinfo {author} {\bibfnamefont {R.~P.}\ \bibnamefont {Muller}},\ }\bibfield
   {title} {\bibinfo {title} {Valley splitting of single-electron si mos
  quantum dots},\ }\href@noop {} {\bibfield  {journal} {\bibinfo  {journal}
  {Appl. Phys. Lett.}\ }\textbf {\bibinfo {volume} {109}},\ \bibinfo {pages}
  {253101} (\bibinfo {year} {2016})}\BibitemShut {NoStop}%
\bibitem [{\citenamefont {Boykin}\ \emph
  {et~al.}(2004{\natexlab{a}})\citenamefont {Boykin}, \citenamefont {Klimeck},
  \citenamefont {Eriksson}, \citenamefont {Friesen}, \citenamefont
  {Coppersmith}, \citenamefont {von Allmen}, \citenamefont {Oyafuso},\ and\
  \citenamefont {Lee}}]{Boykin:2004p115}%
  \BibitemOpen
  \bibfield  {author} {\bibinfo {author} {\bibfnamefont {T.~B.}\ \bibnamefont
  {Boykin}}, \bibinfo {author} {\bibfnamefont {G.}~\bibnamefont {Klimeck}},
  \bibinfo {author} {\bibfnamefont {M.~A.}\ \bibnamefont {Eriksson}}, \bibinfo
  {author} {\bibfnamefont {M.}~\bibnamefont {Friesen}}, \bibinfo {author}
  {\bibfnamefont {S.~N.}\ \bibnamefont {Coppersmith}}, \bibinfo {author}
  {\bibfnamefont {P.}~\bibnamefont {von Allmen}}, \bibinfo {author}
  {\bibfnamefont {F.}~\bibnamefont {Oyafuso}},\ and\ \bibinfo {author}
  {\bibfnamefont {S.}~\bibnamefont {Lee}},\ }\bibfield  {title} {\bibinfo
  {title} {Valley splitting in strained silicon quantum wells},\ }\href
  {https://doi.org/10.1063/1.1637718} {\bibfield  {journal} {\bibinfo
  {journal} {Appl. Phys. Lett.}\ }\textbf {\bibinfo {volume} {84}},\ \bibinfo
  {pages} {115} (\bibinfo {year} {2004}{\natexlab{a}})}\BibitemShut {NoStop}%
\bibitem [{\citenamefont {Boykin}\ \emph
  {et~al.}(2004{\natexlab{b}})\citenamefont {Boykin}, \citenamefont {Klimeck},
  \citenamefont {Friesen}, \citenamefont {Coppersmith}, \citenamefont
  {vonAllmen}, \citenamefont {Oyafuso},\ and\ \citenamefont
  {Lee}}]{Boykin:2004p165325}%
  \BibitemOpen
  \bibfield  {author} {\bibinfo {author} {\bibfnamefont {T.~B.}\ \bibnamefont
  {Boykin}}, \bibinfo {author} {\bibfnamefont {G.}~\bibnamefont {Klimeck}},
  \bibinfo {author} {\bibfnamefont {M.}~\bibnamefont {Friesen}}, \bibinfo
  {author} {\bibfnamefont {S.~N.}\ \bibnamefont {Coppersmith}}, \bibinfo
  {author} {\bibfnamefont {P.}~\bibnamefont {vonAllmen}}, \bibinfo {author}
  {\bibfnamefont {F.}~\bibnamefont {Oyafuso}},\ and\ \bibinfo {author}
  {\bibfnamefont {S.}~\bibnamefont {Lee}},\ }\bibfield  {title} {\bibinfo
  {title} {Valley splitting in low-density quantum-confined heterostructures
  studied using tight-binding models},\ }\href@noop {} {\bibfield  {journal}
  {\bibinfo  {journal} {Phys. Rev. B}\ }\textbf {\bibinfo {volume} {70}},\
  \bibinfo {pages} {165325} (\bibinfo {year} {2004}{\natexlab{b}})}\BibitemShut
  {NoStop}%
\bibitem [{\citenamefont {McJunkin}\ \emph
  {et~al.}(2021{\natexlab{a}})\citenamefont {McJunkin}, \citenamefont
  {MacQuarrie}, \citenamefont {Tom}, \citenamefont {Neyens}, \citenamefont
  {Dodson}, \citenamefont {Thorgrimsson}, \citenamefont {Corrigan},
  \citenamefont {Ercan}, \citenamefont {Savage}, \citenamefont {Lagally},
  \citenamefont {Joynt}, \citenamefont {Coppersmith}, \citenamefont {Friesen},\
  and\ \citenamefont {Eriksson}}]{mcjunkin2021spike}%
  \BibitemOpen
  \bibfield  {author} {\bibinfo {author} {\bibfnamefont {T.}~\bibnamefont
  {McJunkin}}, \bibinfo {author} {\bibfnamefont {E.}~\bibnamefont
  {MacQuarrie}}, \bibinfo {author} {\bibfnamefont {L.}~\bibnamefont {Tom}},
  \bibinfo {author} {\bibfnamefont {S.}~\bibnamefont {Neyens}}, \bibinfo
  {author} {\bibfnamefont {J.}~\bibnamefont {Dodson}}, \bibinfo {author}
  {\bibfnamefont {B.}~\bibnamefont {Thorgrimsson}}, \bibinfo {author}
  {\bibfnamefont {J.}~\bibnamefont {Corrigan}}, \bibinfo {author}
  {\bibfnamefont {H.}~\bibnamefont {Ercan}}, \bibinfo {author} {\bibfnamefont
  {D.}~\bibnamefont {Savage}}, \bibinfo {author} {\bibfnamefont
  {M.}~\bibnamefont {Lagally}}, \bibinfo {author} {\bibfnamefont
  {R.}~\bibnamefont {Joynt}}, \bibinfo {author} {\bibfnamefont
  {S.}~\bibnamefont {Coppersmith}}, \bibinfo {author} {\bibfnamefont
  {M.}~\bibnamefont {Friesen}},\ and\ \bibinfo {author} {\bibfnamefont
  {M.}~\bibnamefont {Eriksson}},\ }\bibfield  {title} {\bibinfo {title} {Valley
  splittings in {S}i/{S}i{G}e quantum dots with a germanium spike in the
  silicon well},\ }\href@noop {} {\bibfield  {journal} {\bibinfo  {journal}
  {Phys. Rev .B}\ }\textbf {\bibinfo {volume} {104}},\ \bibinfo {pages}
  {085406} (\bibinfo {year} {2021}{\natexlab{a}})}\BibitemShut {NoStop}%
\bibitem [{\citenamefont {Wuetz}\ \emph {et~al.}(2021)\citenamefont {Wuetz},
  \citenamefont {Losert}, \citenamefont {Koelliing}, \citenamefont {Stehower},
  \citenamefont {Zwerver}, \citenamefont {Philips}, \citenamefont {Madzik},
  \citenamefont {Xue}, \citenamefont {Zheng}, \citenamefont {Lodari},
  \citenamefont {Amitonov}, \citenamefont {Samkharadze}, \citenamefont
  {Sammak}, \citenamefont {Vandersypen}, \citenamefont {Rahman}, \citenamefont
  {Coppersmith}, \citenamefont {Moutabhir}, \citenamefont {Friesen},\ and\
  \citenamefont {Scappucci}}]{wuetz}%
  \BibitemOpen
  \bibfield  {author} {\bibinfo {author} {\bibfnamefont {B.}~\bibnamefont
  {Wuetz}}, \bibinfo {author} {\bibfnamefont {M.}~\bibnamefont {Losert}},
  \bibinfo {author} {\bibfnamefont {S.}~\bibnamefont {Koelliing}}, \bibinfo
  {author} {\bibfnamefont {L.}~\bibnamefont {Stehower}}, \bibinfo {author}
  {\bibfnamefont {A.}~\bibnamefont {Zwerver}}, \bibinfo {author} {\bibfnamefont
  {S.}~\bibnamefont {Philips}}, \bibinfo {author} {\bibfnamefont
  {M.}~\bibnamefont {Madzik}}, \bibinfo {author} {\bibfnamefont
  {X.}~\bibnamefont {Xue}}, \bibinfo {author} {\bibfnamefont {G.}~\bibnamefont
  {Zheng}}, \bibinfo {author} {\bibfnamefont {M.}~\bibnamefont {Lodari}},
  \bibinfo {author} {\bibfnamefont {S.}~\bibnamefont {Amitonov}}, \bibinfo
  {author} {\bibfnamefont {N.}~\bibnamefont {Samkharadze}}, \bibinfo {author}
  {\bibfnamefont {A.}~\bibnamefont {Sammak}}, \bibinfo {author} {\bibfnamefont
  {L.}~\bibnamefont {Vandersypen}}, \bibinfo {author} {\bibfnamefont
  {R.}~\bibnamefont {Rahman}}, \bibinfo {author} {\bibfnamefont
  {S.}~\bibnamefont {Coppersmith}}, \bibinfo {author} {\bibfnamefont
  {O.}~\bibnamefont {Moutabhir}}, \bibinfo {author} {\bibfnamefont
  {M.}~\bibnamefont {Friesen}},\ and\ \bibinfo {author} {\bibfnamefont
  {G.}~\bibnamefont {Scappucci}},\ }\bibfield  {title} {\bibinfo {title}
  {Atomic fluctuations lifting the energy degeneracy in {S}i/{S}i{G}e quantum
  dots},\ }\href@noop {} {\bibfield  {journal} {\bibinfo  {journal} {arXiv
  preprint ArXiv:2112.09606}\ } (\bibinfo {year} {2021})}\BibitemShut {NoStop}%
\bibitem [{\citenamefont {McJunkin}\ \emph
  {et~al.}(2021{\natexlab{b}})\citenamefont {McJunkin}, \citenamefont {Harpt},
  \citenamefont {Feng}, \citenamefont {Losert}, \citenamefont {Rahman},
  \citenamefont {Dodson}, \citenamefont {Wolff}, \citenamefont {Savage},
  \citenamefont {Lagally}, \citenamefont {Joynt}, \citenamefont {Coppersmith},
  \citenamefont {Friesen},\ and\ \citenamefont
  {Eriksson}}]{mcjunkin2021wigglewell}%
  \BibitemOpen
  \bibfield  {author} {\bibinfo {author} {\bibfnamefont {T.}~\bibnamefont
  {McJunkin}}, \bibinfo {author} {\bibfnamefont {B.}~\bibnamefont {Harpt}},
  \bibinfo {author} {\bibfnamefont {Y.}~\bibnamefont {Feng}}, \bibinfo {author}
  {\bibfnamefont {M.}~\bibnamefont {Losert}}, \bibinfo {author} {\bibfnamefont
  {R.}~\bibnamefont {Rahman}}, \bibinfo {author} {\bibfnamefont
  {J.}~\bibnamefont {Dodson}}, \bibinfo {author} {\bibfnamefont
  {M.}~\bibnamefont {Wolff}}, \bibinfo {author} {\bibfnamefont
  {D.}~\bibnamefont {Savage}}, \bibinfo {author} {\bibfnamefont
  {M.}~\bibnamefont {Lagally}}, \bibinfo {author} {\bibfnamefont
  {R.}~\bibnamefont {Joynt}}, \bibinfo {author} {\bibfnamefont
  {S.}~\bibnamefont {Coppersmith}}, \bibinfo {author} {\bibfnamefont
  {M.}~\bibnamefont {Friesen}},\ and\ \bibinfo {author} {\bibfnamefont
  {M.}~\bibnamefont {Eriksson}},\ }\bibfield  {title} {\bibinfo {title}
  {{S}i{G}e quantum wells with oscillating {G}e concentrations for quantum dot
  qubits},\ }\href@noop {} {\bibfield  {journal} {\bibinfo  {journal} {arXiv
  preprint arXiv:2112.09765}\ } (\bibinfo {year}
  {2021}{\natexlab{b}})}\BibitemShut {NoStop}%
\bibitem [{\citenamefont {Ando}\ \emph {et~al.}(1982)\citenamefont {Ando},
  \citenamefont {Fowler},\ and\ \citenamefont {Stern}}]{Ando:1982p437}%
  \BibitemOpen
  \bibfield  {author} {\bibinfo {author} {\bibfnamefont {T.}~\bibnamefont
  {Ando}}, \bibinfo {author} {\bibfnamefont {A.~B.}\ \bibnamefont {Fowler}},\
  and\ \bibinfo {author} {\bibfnamefont {F.}~\bibnamefont {Stern}},\ }\bibfield
   {title} {\bibinfo {title} {Electronic properties of two-dimensional
  systems},\ }\href@noop {} {\bibfield  {journal} {\bibinfo  {journal} {Rev.
  Mod. Phys.}\ }\textbf {\bibinfo {volume} {54}},\ \bibinfo {pages} {437}
  (\bibinfo {year} {1982})}\BibitemShut {NoStop}%
\bibitem [{\citenamefont {Sch\"{a}ffler}(1997)}]{schaeffler1997}%
  \BibitemOpen
  \bibfield  {author} {\bibinfo {author} {\bibfnamefont {F.}~\bibnamefont
  {Sch\"{a}ffler}},\ }\bibfield  {title} {\bibinfo {title} {High-mobility {S}i
  and {G}e structures},\ }\href@noop {} {\bibfield  {journal} {\bibinfo
  {journal} {Semicon. Sci. Tech.}\ }\textbf {\bibinfo {volume} {12}},\ \bibinfo
  {pages} {1515} (\bibinfo {year} {1997})}\BibitemShut {NoStop}%
\bibitem [{\citenamefont {Saraiva}\ \emph {et~al.}(2011)\citenamefont
  {Saraiva}, \citenamefont {Calderon}, \citenamefont {Capaz}, \citenamefont
  {Hu}, \citenamefont {Sarma},\ and\ \citenamefont {Koiller.}}]{saraiva}%
  \BibitemOpen
  \bibfield  {author} {\bibinfo {author} {\bibfnamefont {A.~L.}\ \bibnamefont
  {Saraiva}}, \bibinfo {author} {\bibfnamefont {M.~J.}\ \bibnamefont
  {Calderon}}, \bibinfo {author} {\bibfnamefont {R.~B.}\ \bibnamefont {Capaz}},
  \bibinfo {author} {\bibfnamefont {X.}~\bibnamefont {Hu}}, \bibinfo {author}
  {\bibfnamefont {S.~D.}\ \bibnamefont {Sarma}},\ and\ \bibinfo {author}
  {\bibfnamefont {B.}~\bibnamefont {Koiller.}},\ }\bibfield  {title} {\bibinfo
  {title} {Intervalley coupling for interface-bound electrons in silicon: An
  effective mass study},\ }\href@noop {} {\bibfield  {journal} {\bibinfo
  {journal} {Physical Review B}\ }\textbf {\bibinfo {volume} {84}},\ \bibinfo
  {pages} {155320} (\bibinfo {year} {2011})}\BibitemShut {NoStop}%
\bibitem [{\citenamefont {Goswami}\ \emph
  {et~al.}(2007{\natexlab{b}})\citenamefont {Goswami}, \citenamefont {Slinker},
  \citenamefont {Friesen}, \citenamefont {McGuire}, \citenamefont {Truitt},
  \citenamefont {Tahan}, \citenamefont {Klein}, \citenamefont {Chu},
  \citenamefont {Mooney}, \citenamefont {van~der Weide}, \citenamefont {Joynt},
  \citenamefont {Coppersmith},\ and\ \citenamefont {Eriksson}}]{goswami}%
  \BibitemOpen
  \bibfield  {author} {\bibinfo {author} {\bibfnamefont {S.}~\bibnamefont
  {Goswami}}, \bibinfo {author} {\bibfnamefont {K.}~\bibnamefont {Slinker}},
  \bibinfo {author} {\bibfnamefont {M.}~\bibnamefont {Friesen}}, \bibinfo
  {author} {\bibfnamefont {L.}~\bibnamefont {McGuire}}, \bibinfo {author}
  {\bibfnamefont {J.}~\bibnamefont {Truitt}}, \bibinfo {author} {\bibfnamefont
  {C.}~\bibnamefont {Tahan}}, \bibinfo {author} {\bibfnamefont
  {L.}~\bibnamefont {Klein}}, \bibinfo {author} {\bibfnamefont
  {J.}~\bibnamefont {Chu}}, \bibinfo {author} {\bibfnamefont {P.}~\bibnamefont
  {Mooney}}, \bibinfo {author} {\bibfnamefont {D.}~\bibnamefont {van~der
  Weide}}, \bibinfo {author} {\bibfnamefont {R.}~\bibnamefont {Joynt}},
  \bibinfo {author} {\bibfnamefont {S.}~\bibnamefont {Coppersmith}},\ and\
  \bibinfo {author} {\bibfnamefont {M.}~\bibnamefont {Eriksson}},\ }\bibfield
  {title} {\bibinfo {title} {Controllable valley splitting in silicon quantum
  devices},\ }\href@noop {} {\bibfield  {journal} {\bibinfo  {journal} {Nature
  Physics}\ }\textbf {\bibinfo {volume} {3}},\ \bibinfo {pages} {41} (\bibinfo
  {year} {2007}{\natexlab{b}})}\BibitemShut {NoStop}%
\bibitem [{\citenamefont {Chelikowski}\ \emph {et~al.}(1973)\citenamefont
  {Chelikowski}, \citenamefont {Chadi},\ and\ \citenamefont
  {Cohen}}]{chelikowski1}%
  \BibitemOpen
  \bibfield  {author} {\bibinfo {author} {\bibfnamefont {J.}~\bibnamefont
  {Chelikowski}}, \bibinfo {author} {\bibfnamefont {D.}~\bibnamefont {Chadi}},\
  and\ \bibinfo {author} {\bibfnamefont {M.}~\bibnamefont {Cohen}},\ }\bibfield
   {title} {\bibinfo {title} {Calculated valence-band densities of states and
  photoemission spectra of diamond and zinc-blende semiconductors},\
  }\href@noop {} {\bibfield  {journal} {\bibinfo  {journal} {Phys. Rev. B}\
  }\textbf {\bibinfo {volume} {8}},\ \bibinfo {pages} {085406} (\bibinfo {year}
  {1973})}\BibitemShut {NoStop}%
\bibitem [{\citenamefont {Chelikowski}\ and\ \citenamefont
  {Cohen}(1976)}]{chelikowski2}%
  \BibitemOpen
  \bibfield  {author} {\bibinfo {author} {\bibfnamefont {J.}~\bibnamefont
  {Chelikowski}}\ and\ \bibinfo {author} {\bibfnamefont {M.}~\bibnamefont
  {Cohen}},\ }\bibfield  {title} {\bibinfo {title} {Nonlocal pseudopotential
  calculations for the electronic structure of eleven diamond and zinc-blende
  semiconductors},\ }\href@noop {} {\bibfield  {journal} {\bibinfo  {journal}
  {Phys. Rev. B}\ }\textbf {\bibinfo {volume} {14}},\ \bibinfo {pages} {556}
  (\bibinfo {year} {1976})}\BibitemShut {NoStop}%
\bibitem [{\citenamefont {Kohn}(1957)}]{kohn}%
  \BibitemOpen
  \bibfield  {author} {\bibinfo {author} {\bibfnamefont {W.}~\bibnamefont
  {Kohn}},\ }\bibfield  {title} {\bibinfo {title} {Shallow impurity states in
  silicon and germanium},\ }\href@noop {} {\bibfield  {journal} {\bibinfo
  {journal} {Solid State Physics}\ }\textbf {\bibinfo {volume} {5}},\ \bibinfo
  {pages} {257} (\bibinfo {year} {1957})}\BibitemShut {NoStop}%
\end{thebibliography}%

\end{document}